\documentclass{emulateapj}
\usepackage{times}

\shorttitle{Numerical Galaxy Catalog I.}
\shortauthors{NAGASHIMA ET AL.}

\begin{document}

\title{Numerical Galaxy Catalog -I. A Semi-analytic Model of
Galaxy Formation with $N$-body simulations}

\author{Masahiro Nagashima,\altaffilmark{1,2,3}
Hideki Yahagi,\altaffilmark{3,4}
Motohiro Enoki,\altaffilmark{5,6}
Yuzuru Yoshii,\altaffilmark{7,8}
and
Naoteru Gouda\altaffilmark{5}
}
\email{masa@scphys.kyoto-u.ac.jp}
\altaffiltext{1}{Department of Physics, Graduate School of Science,
Kyoto University, Sakyo-ku, Kyoto 606-8502, Japan}
\altaffiltext{2}{Department of Physics, University of Durham, South Road, Durham DH1 3LE, U.K.}
\altaffiltext{3}{Division of Theoretical Astrophysics, National Astronomical
Observatory, National Institute of Natural Science, Mitaka, Tokyo 181-8588, Japan}
\altaffiltext{4}{Department of Astronomy, University of Tokyo,
Bunkyo-ku, Tokyo 113-0033, Japan}
\altaffiltext{5}{Division of Astrometry and Celestial Mechanics, National
Astronomical Observatory, National Institute of Natural Science, Mitaka, Tokyo 181-8588, Japan}
\altaffiltext{6}{Astronomical Data Analysis Center, National
Astronomical Observatory, National Institute of Natural Science, Mitaka, Tokyo 181-8588, Japan}
\altaffiltext{7}{
Institute of Astronomy, School of Science, The University of
Tokyo, Mitaka, Tokyo 181-0015, Japan}
\altaffiltext{8}{Research Center for the Early Universe, School of Science, The
University of Tokyo, Bunkyo-ku, Tokyo 113-0033, Japan}

\begin{abstract}
We construct the {\it Numerical Galaxy Catalog} ($\nu$GC), based on a
semi-analytic model of galaxy formation combined with high-resolution
$N$-body simulations in a $\Lambda$-dominated flat cold dark matter
($\Lambda$CDM) cosmological model.  The model includes several essential
ingredients for galaxy formation, such as merging histories of dark
halos directly taken from $N$-body simulations, radiative gas cooling,
star formation, heating by supernova explosions (supernova feedback),
mergers of galaxies, population synthesis, and extinction by internal
dust and intervening \ion{H}{1} clouds.  As the first paper in a series
using this model, we focus on basic photometric, structural and
kinematical properties of galaxies at present and high redshifts.  Two
sets of model parameters are examined, strong and weak supernova
feedback models, which are in good agreement with observational
luminosity functions of local galaxies in a range of observational
uncertainty.  Both models agree well with many observations such as cold
gas mass-to-stellar luminosity ratios of spiral galaxies, \ion{H}{1}
mass functions, galaxy sizes, faint galaxy number counts and photometric
redshift distributions in optical pass-bands, isophotal angular sizes,
and cosmic star formation rates.  In particular, the strong supernova
feedback model is in much better agreement with near-infrared
($K'$-band) faint galaxy number counts and redshift distribution than
the weak feedback model and our previous semi-analytic models based on
the extended Press-Schechter formalism.  Observed Tully-Fisher relations
for bright galaxies and color-magnitude relations for dwarf elliptical
galaxies in clusters of galaxies are broadly reproduced, but no
agreement with observations is obtained over whole ranges.
Nevertheless, they are improved compared with results produced by other
semi-analytic models.  A direction for overcoming the remaining problems
is discussed.  We also find that the resolution of $N$-body simulations,
which is down to $3\times 10^{9} M_{\odot}$ for the minimum mass of dark
halos consisting of ten dark matter particles in our model, plays a
significant role in galaxy formation, and that merging histories of dark
halos directly taken from $N$-body simulations produce results different
from models based on the extended Press-Schechter model even if the mass
function of dark halos at present is set to be the same as that obtained
by the same $N$-body simulations used here.
\end{abstract}


\section{Introduction}

Galaxies are one of the most important hierarchies composing the
Universe.  For example, in order to understand the evolution of the
Universe and its structure, galaxies are often used as a measure of
space-time in performing wide-field or deep surveys.  In the field of
observational cosmology, therefore, galaxies are indispensable units in
the Universe.  What we should note about galaxies is, however, that
galaxies have also been evolving along with the evolution of the
Universe.  Thus, correctly modeling the evolution of galaxies is
essential in observational cosmology.  At the same time, galaxies
themselves are also interesting objects because many physical processes
are involved in galaxy formation.  However, since some processes such as
star formation are still poorly understood, it has been challenging to
understand galaxy formation.  To attack this subject, correctly modeling
galaxy formation is critical as well, particularly under the current
situation that numerical hydrodynamical simulations of galaxy formation
still have large uncertainties \citep[see, e.g.,][]{okamoto03,
okamoto05}.  Thus, the purpose of this paper is to construct a better
and more reliable model of galaxy formation aided by high resolution
$N$-body simulations, which are well established in contrast to
hydrodynamical simulations of galaxy formation, with quality high enough
to be compared with recent high precision observations.

It has been widely understood that the cold dark matter (CDM) model well
describes the Universe.  Recent high resolution observations such as
those measuring anisotropies of the cosmic microwave background by the
{\it Wilkinson Microwave Anisotropy Probe} \citep{s03} and the type Ia
supernova (SN) rate \citep[e.g.][]{riess98, perlmutter99} have revealed
that the energy density in the Universe is dominated by dark energy or
the cosmological constant, and that gravity is dominated by the CDM.
The large-scale structure of the universe has evolved from small density
fluctuations existing in the early universe via gravitational
instability.  Because the predicted amplitude of density fluctuations is
monotonically increasing toward smaller scales, small-mass dark halos,
which are virialized objects of dark matter, firstly collapse and then
those of larger masses subsequently collapse swallowing the smaller dark
halos.  Therefore, the formation of cosmological structure predicted by
the CDM model is often called the hierarchical clustering.  Most
baryonic processes, such as gas cooling and star formation, take place
within dark halos because of the gravitational potential of dark matter.
Therefore, galaxy formation models must be constructed within the framework of
hierarchical clustering as a natural consequence of the CDM model.

Along this line, semi-analytic (SA) modeling of galaxy formation has
been developed \citep[e.g.][]{kwg93, cafnz94, clbf00, vankampen99, sp99,
ngs99, ntgy01, nytg02, mcfgp02, hatton03}.  In most SA models, merging
histories or {\it merger trees} of dark halos are realized by using a
Monte Carlo method based on extended Press-Schechter (EPS) models
\citep{ph90, bcek91, b91, lc93, kw93, sk99, clbf00} which provide mass
functions of progenitor halos.  The EPS formalism is an extension of the
Press-Schechter (PS) model of mass functions of dark halos \citep{ps74},
which are derived from the power spectrum of initial density
fluctuations combined with a spherically symmetric collapse model to
describe non-linear evolution of density fluctuations \citep{t69, gg72}.
Considering density fluctuations within an overdense region which
collapses later, mass functions of progenitor halos of a larger, later
collapsing halo are obtained.  SA models also include several important
physical processes such as gas cooling, star formation, heating by SN
explosions (SN feedback) and galaxy mergers, which take place within
dark halos.  Moreover, to directly compare theoretical results with
observations, a stellar population synthesis technique is usually
incorporated.  Because we do not follow the dynamics related with these
processes as hydrodynamical simulations, the computation is much faster.
This means that we can explore a much larger space of parameters
characterizing these processes, and therefore SA modeling is an
important approach, complementary to hydrodynamical simulations in
understanding the physics of galaxy formation.

In addition to those galaxy formation processes, observations of high
redshift galaxies suffer from several effects that are not intrinsic in
galaxies.  \citet{bcfl98} and \citet{ntgy01, nytg02} considered
absorption of emitted light from high redshift galaxies by intervening
\ion{H}{1} clouds \citep{yp94, madau95}.  The latter work also took into
account selection effects arising from cosmological dimming of surface
brightness dependent on observations \citep{y93, ty00, t01}.

Furthermore, a new ingredient related to SN feedback has recently been
introduced into an SA model by \citet[][hereafter NY04]{ny04} for the
first time.  They considered dynamical response to starburst-induced gas
removal from galaxies on the size and velocity dispersion
\citep[e.g.][]{ya87}, followed by a change in the depth of the
gravitational potential well.  They also showed that the model can
successfully explain many aspects of photometric, structural and
kinematical properties of dwarf ellipticals.  This is called the Mitaka
model, and is the basis of the SA model we shall show in this paper.

The PS model, however, has been recognized to be consistent with
$N$-body simulations only within a factor of two \citep{yng96, ng97,
monaco98, n01, wht02, reed03}.  Some SA models, therefore, use corrected
mass functions based on high resolution $N$-body simulations
\citep{st99, j01, yny04} instead of the PS model.  \citet{bcfbl00,
bfbcl01} used information on halo distribution from an $N$-body
simulation and adapted their SA models, GALFORM \citep{clbf00}, in which
an EPS model is used.  This method has the virtue, not only of allowing
spatial distribution of galaxies to be followed, but also of resolution
to low-mass halos at high redshift being freely set from the $N$-body
simulations.  Unfortunately, we cannot avoid the limitations of the PS
model with these approaches, because the problems of EPS formalism still
remain.  To bypass the limitations of EPS formalism, hybrid SA models
have been developed in which merger trees of dark halos directly taken
from $N$-body simulations are used, while we have to carefully interpret
any results from such hybrid models because of the limited numerical
resolution.  Fortunately, currently available $N$-body simulations are
of high enough resolution to resolve the effective Jeans mass after the
cosmic reionization.  This kind of hybrid model can also be used to
derive spatial distributions of galaxies.  Thus, combined with
photometric properties, mock catalogs can be made for comparison with
individual galaxy surveys, without any loss of SA model advantages.
Therefore, this approach provides strong tools with which to study
galaxy formation.

The pioneering work for this approach was carried out by \citet{rpqr97}.
Since then, their work has been substantially improved upon by
\citet{kcdw99a,kcdw99b}, \citet{dkcw99, dkbwse01}, \citet{slsdkw01},
\citet{hcfbbl03, hcfbblp03}, and \citet{hatton03}.  We have learned many
things about the spatial clustering of galaxies, particularly of high
redshift galaxies, from those models.  However, as we shall show in this
paper, those models suffer from limited numerical resolution of $N$-body
simulations.  As another direction to the extension of SA models with
$N$-body simulations, we should note works by \citet{on01, on03} and
\citet{swtk01}, in which merging histories of {\it subhalos} in galaxy
clusters are taken into account.  This enables us to derive
relationships between galaxy properties and spatial distribution in
halos, such as the morphology-density relation \citep[e.g.][]{d80}.  In
this paper, we do not follow the merging histories of subhalos, but it
would be worth including those in future.

In this paper, based on our previous work on the Mitaka SA model (NY04)
and $N$-body simulations using a parallel adaptive mesh refinement (AMR)
code \citep{y05}, we construct the {\it Numerical Galaxy Catalog}
($\nu$GC), which is an SA model combined with high resolution $N$-body
simulations.  As shown below, the mass resolution is high enough to
resolve the Jeans mass, and the size of the simulation box is also large
enough to express a fair sample of the universe.  We refer the reader to
\citet{yny04}, in which mass functions of dark halos are presented, to
see how powerful this AMR $N$-body code is.  As a first paper in a
series of $\nu$GC, here we focus on photometric, structural and
kinematical properties of galaxies at present and high redshift such as
luminosity functions, cold gas fractions, sizes, the Tully-Fisher
relation, faint galaxy number counts, redshift distributions, cosmic
star formation histories, and so on.  Spatial distribution of galaxies
is discussed in a subsequent paper \citep{ynegy}.  We also plan to
extend this model in future \citep{eyngy} to include a quasar formation
model developed by \citet{eng03} and \citet{eins04}.

This paper is outlined as follows: In \S2 we describe our SA model.  In
\S3 we summarize model parameters in our SA model.  In \S4 we compare
the theoretical predictions of SA model galaxies with various
observations at present.  In \S5 we examine high redshift galaxies
predicted by our SA model with observations.  In \S6 we provide summary
and conclusions.

\section{Model}
The galaxy formation scenario on which our model is based is as follows:
In the CDM universe, small dark matter halos compared with typical
present mass of halos emerge through non-linear gravitational evolution
of small density fluctuations.  Such dark halos cluster and merge into
larger halos in a manner that depends on the adopted power spectrum of
the initial density fluctuations.  In each of the merged dark halos,
radiative cooling of virialized hot gas, star formation, and gas
reheating by supernovae occur.  The cooled dense gas and stars
constitute {\it galaxies}.  Dark halos are filled with the virialized
hot gas, so if the mass of halos is sufficiently large, it should be
recognized as an intracluster medium.  These galaxies sometimes merge in
a common dark halo, and then more massive galaxies form.  Repeating
these processes, galaxies form and evolve to the present epoch.

Merging histories of dark halos are directly taken from $N$-body
simulations \citep{yny04}.  The galaxy formation model running on the
merging histories of dark halos is almost the same as the Mitaka model
given by NY04, in which the merging histories were realized by using a
Monte Carlo method based on the EPS formalism provided by \citet{sk99}.

Some ingredients of our SA model are revised from those in NY04.
Modifications include a model of the star formation (SF) time-scale, a
prescription for mergers between satellite galaxies, and a dust
extinction model during starbursts.  The details are described below.

Throughout this paper, a recent standard $\Lambda$CDM model is
considered, that is, $\Omega_{0}=0.3, \Omega_{\Lambda}=0.7, h=0.7$ and
$\sigma_{8}=0.9$, where these parameters denote the mean density of the
universe, the cosmological constant, the Hubble parameter and the
normalization of the power spectrum of the initial density fluctuation
field.  The shape of the power spectrum given by \citet{s95} is used, an
extension of \citet{bbks} taking into account the effects of baryons.
The cosmic baryon density $\Omega_{\rm b}=0.048$ is adopted \citep{s03}.
Those parameters are tabulated in Table \ref{tab:cosmo}.  Model
parameters described below are summarized in \S 3.

\begin{deluxetable}{ccccc}
\tabletypesize{\scriptsize}
\tablecaption{Cosmological Parameters\label{tab:cosmo}}  
\tablewidth{0pt}
\tablehead{
\colhead{$\Omega_{0}$} & \colhead{$\Omega_{\Lambda}$} & \colhead{$h$} &
 \colhead{$\sigma_{8}$} & \colhead{$\Omega_{\rm b}$}}
\startdata
0.3 & 0.7 & 0.7 & 0.9 & 0.048
\enddata

\tablecomments{Cosmological parameters are based on the {\it WMAP}
results \citep{s03}.}
\end{deluxetable}

\subsection{$N$-body Simulations and Construction of Merging Histories of Dark Halos}\label{sec:mh}

Merger trees of dark halos are directly taken from cosmological $N$-body
simulations.  Our simulation has $N=512^{3}$ dark matter particles in a
box size of $L=70h^{-1}$Mpc.  Since the minimum number of particles
identifying a dark halo is 10, the minimum mass of dark halos $M_{70}$
is nearly equal to $3.04\times 10^{9}M_{\odot}$.  The simulation code is
developed by \citet{y05}, based on AMR technique to compute the
gravitational force \citep{krakk97, knegb01, yy01, tey02} with periodic
boundary conditions.  The method of identifying dark halos is the
friends-of-friends (FoF) grouping algorithm \citep{defw}, with the
linking length $b=0.2$ as usual.

\begin{figure}
\plotone{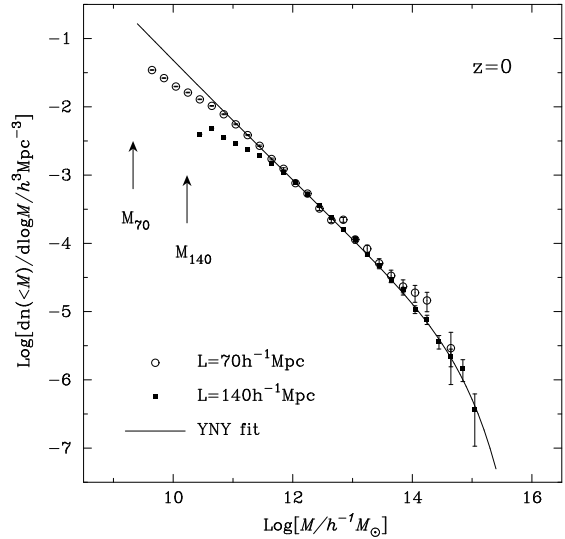}

\caption{Mass functions of dark halos at $z=0$.  The open circles and
solid squares with error bars represent mass functions given by the
simulations of $L=70 h^{-1}$ Mpc and 140 $h^{-1}$ Mpc, respectively.
The error bars indicate 1 $\sigma$ scatter.  The minimum masses of those
simulations are indicated by $M_{70}$ and $M_{140}$ in the figure.  The
solid line is a fitting function to results of several simulations
\citep{yny04}.  } \label{fig:mf}
\end{figure}

Rigorously speaking, however, fully resolved mass might be larger, about
$5\times 10^{11}M_{\odot}$, at which the resultant mass function
deviates from a power law, as indicated in Figure \ref{fig:mf} by open
circles with error bars.  The solid line in the figure is a fitting
function for several simulations performed with different box sizes from
35 to 140 $h^{-1}$ Mpc proposed by \citet{yny04}.  To evaluate the
resolution effects on galaxy formation, we shall use another set of
$N$-body results of a 140 $h^{-1}$ Mpc box with the same number of
particles, $N=512^{3}$ in \S\S\ref{sec:lf}.  The mass function derived
by this simulation is indicated in the figure by solid squares, with a
minimum mass of $M_{140}=8M_{70}$.  Parameters of those two simulations,
``70'' and ``140'', are tabulated in Table \ref{tab:nbody}.

\begin{deluxetable}{ccccc}
\tabletypesize{\scriptsize}
\tablecaption{Parameters of $N$-body simulations\label{tab:nbody}}  
\tablewidth{0pt}
\tablehead{
\colhead{Simulation} & \colhead{$L$ ($h^{-1}$Mpc)} & \colhead{$N$}
 & \colhead{Minimum Halo Mass ($M_{\odot}$)} & \colhead{$b$} }
\startdata
70 & 70 & 512$^{3}$ & $3.04\times 10^{9} (M_{70})$ & 0.2\\
140 & 140 & 512$^{3}$ & $2.43\times 10^{10} (M_{140})$ & 0.2
\enddata

\tablecomments{$L$ is the size of the simulation box, $N$ is the number
 of dark matter particles, and $b$ is the linking length parameter for
 identifying isolated dark halos using the FOF algorithm.  Particles
 within $b$ times a mean separation of particles are grouped.  The
 minimum halos are defined as those of 10 particles.}
\end{deluxetable}

\begin{figure}
\plotone{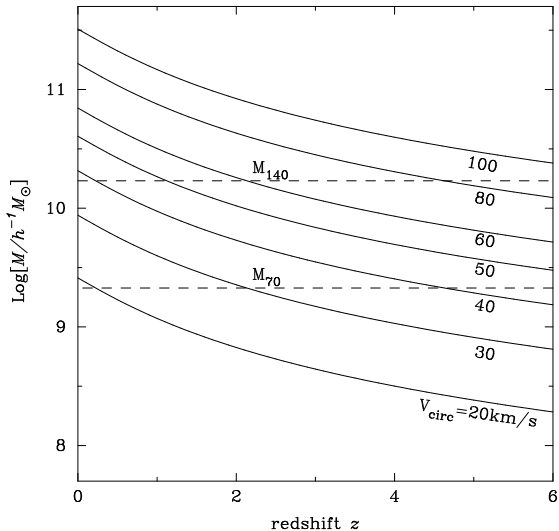}

\caption{Relationship between mass and circular velocity of dark halos
as a function of redshift.  The solid lines indicate the mass of halos
with constant circular velocities, $V_{\rm circ}=20, ... 100$ km
s$^{-1}$, from the bottom to the top, as indicated in the figure.  The
dashed lines represent the minimum masses $M_{70}$ and $M_{140}$ of the
simulations with $70$ and $140 h^{-1}$ Mpc box sizes, respectively.  }

\label{fig:vm}
\end{figure}

Figure \ref{fig:vm} shows the importance of examining the mass
resolution of $N$-body simulations on galaxy formation.  The solid lines
indicate masses of dark halos for constant circular velocities, $V_{\rm
circ}=$20, 30, 40, 50, 60, 80, and 100 km s$^{-1}$ as a function of
redshift from the bottom to the top.  We use the spherical collapse
model to obtain the relationship between the mass and circular velocity,
which is expressed as
\begin{equation}
 M=3.26\times 10^{6}\Delta(z)^{-1/2}\left(\frac{V_{\rm circ}}{\rm km~ s^{-1}}\right)^{3} h^{-1} M_{\odot},
\end{equation}
where $\Delta(z)$ is the ratio of the mean density in halos collapsing
at $z$ to the critical density of the universe.  The dashed lines are
the minimum masses $M_{70}$ and $M_{140}$ as indicated in the figure.

After the cosmic reionization, the Jeans mass should be at least $V_{\rm
circ}\simeq 20$ km s$^{-1}$, which corresponds to $T\simeq 10^{4}$ K.
Based on his hydrodynamical simulations, \citet{gnedin00} suggested that
an effective Jeans mass, or the {\it filtering mass}, becomes even
larger affected by the past thermal history, although the results for
low redshifts were an extrapolation aided by a linear perturbation
theory.  \citet{somerville02} has reported the following result on the
effective Jeans mass: the mass of halos which retain half of the gas is
well approximated by mass corresponding to circular velocity $V_{\rm
circ}\simeq 50$ km~s$^{-1}$; halos with $V_{\rm circ}<30$ km~s$^{-1}$
are not able to accrete most of the gas and those with $V_{\rm circ}>75$
km~s$^{-1}$ are almost unaffected.  As shown in Figure \ref{fig:vm},
$M_{70}$ corresponds to $V_{\rm circ}\simeq 20$ km s$^{-1}$ at $z\simeq
0$ and 40 km s$^{-1}$ at $z\simeq 5$.  These values should be sufficient
to resolve the effective Jeans mass.  On the other hand, $M_{140}$
corresponds to $V_{\rm circ}\simeq 40$ km s$^{-1}$ at $z\simeq 0$ and 80
km s$^{-1}$ at $z\simeq 5$.  These are marginal, unless the effective
Jeans mass is similar to or larger than that estimated by
\citet{gnedin00}.  Thus our 70 $h^{-1}$Mpc simulation is quite plausible
for correct modeling of galaxy formation.  Apart from detailed
discussion on the Jeans mass, we shall study the effects of the
resolution on luminosity functions in \S\S \ref{sec:lf}.  Note that
other theoretical data in the public domain provided by SA models
combined with $N$-body simulations have much larger masses of dark
matter particles.  For example, the particle mass in the GIF project
\citep{kcdw99a} is about 65 times larger than ours, and that of the
GalICS project \citep{hatton03} is about 27 times larger.  These masses
are even larger than our 140 $h^{-1}$ Mpc box simulation.

In our simulation, first collapsing halos emerge at $z\simeq 18$.  Then
we stock data sets of the particle distribution for approximately 70
time slices.  When identifying dark halos, a marker particle for each
halo is also defined, which is a particle whose mechanical energy is
minimum.  The marker particle is assigned to the halo's central galaxy,
which is defined below.  The position of galaxies is specified using the
corresponding marker particles.  A halo at time $t_{i}$ is connected
with another halo at the next time-step $t_{i+1} (> t_{i})$ as its
progenitor halo, which has the largest number of dark matter particles
common to the both.

Some collapsed halos, especially small halos, disappear later.  Such
``floating'' halos should be fake due to fluctuations of particle
distribution, so we remove the floating halos from the sets of merger
trees.  Some collapsed halos sometimes fragment, particularly at low
redshift.  This is caused by highly dense clumps penetrating their host
halos, which go back into and are absorbed by the host halos again
later.  To avoid such a ``loop'' structure in merging histories, the
fragmented halos are forcedly merged into the host halos.  Then each
merger tree has a {\it tree structure} with a {\it root} halo at $z=0$.
Almost all fragmented halos are small and less than 10\% of their host
halos in mass.  Because the resolution of our $N$-body simulation is
very high, small halos swallowed by larger halos likely survive and
penetrate.  Physically, gas in such clumps should not fragment and
should quickly merge into hot gas in their host halos, apart from dark
matter, because gas is collisional different from collisionless dark
matter.  Therefore, as far as we focus on the evolution of baryons, this
manipulation should be performed.  The details of this simulation can be
found in \citet{y05} and \citet{yny04}.

\subsection{Tidal Stripping of Subhalos}\label{sec:tidal}
Recent high resolution $N$-body simulations of galaxy clusters have
revealed that swallowed dark halos survive in their host halos as {\it
subhalos}.  Envelopes of these subhalos are stripped by tidal force from
the host halo.  In our simulations, such subhalos are not identified but
are taken into account by the following model.  We assume the radius of
a tidally stripped subhalo $r_{\rm t}$, which had a radius of $r_{\rm
s}$ and mass of $M_{\rm s}$ before the stripping, by
\begin{equation}
 \frac{r_{\rm t}}{r_{\rm s}}=\frac{r_{\rm peri}}{r_{\rm apo}}\frac{M_{\rm h}}{M_{\rm s}}\left(\frac{V_{\rm circ,s}}{V_{\rm 
circ,h}}\right)^{3},
\end{equation}
where $r_{\rm peri}$ and $r_{\rm apo}$ are respectively the pericenter
and apocenter for the orbit of the subhalo, and subscripts ``h'' and
``s'' indicate the host halo and subhalo.  In this paper, a ratio of
$r_{\rm peri}/r_{\rm apo}=0.2$ is assumed, as in NY04 \citep{oh99,
oh00}.  Now we assume a singular isothermal distribution for dark matter
in subhalos.  This means that the decrease in mass by the tidal
stripping is proportional to $r_{\rm t}/r_{\rm s}$.  Thus, the tidal
stripping significantly affects the time-scale of galaxy mergers (\S\S
\ref{sec:merger}).

\subsection{Hot Halo Gas and Its Cooling}
Firstly, we define the formation of halos as follows: Dark halos grow as
time passes via mergers and accretion of other halos.  Hereafter, the
formation epoch of halos is referred to as the time when the mass of a
dark halo exceeds twice the mass of the halo at the final formation
epoch \citep{lc93}.  At this time, the circular velocity of the halo is
reestimated, since the halo is regarded as collapsing at this epoch.
The mean mass density in dark halos is assumed to be proportional to the
cosmic mean density at the formation epoch following the spherically
symmetric collapse model, independent of the $N$-body results.  Each
collapsing dark halo contains baryonic matter with a mass fraction
$\Omega_{\rm b}/\Omega_{0}$.  The baryonic matter consists of diffuse
hot gas, dense cold gas, and stars.

When a halo  forms with a circular velocity $V_{\rm circ}$, the hot
gas contained in the halo is shock-heated to the virial temperature of
the halo,
\begin{equation}
 T_{\rm vir}=\frac{1}{2}\frac{\mu m_{\rm p}}{k_{\rm B}}V_{\rm circ}^{2},
\end{equation}
where $m_{\rm p}, k_{\rm B}$ and $\mu$ are the proton mass, the
Boltzmann constant, and the mean molecular weight.  The hot gas density
is assumed to have an isothermal density profile with a finite core
radius, according to \citet{skss02},
\begin{equation}
 \rho_{\rm hot}(r)=\frac{\rho_{\rm hot,0}}{1+(r/r_{c})^{2}},
\end{equation} 
where $r_{c}=0.22R_{\rm vir}/c$, $R_{\rm vir}$ is the virial radius of
the host halo, and $c$ is the concentration modeled by \citet{bksskkpd01},
\begin{equation}
 c(M,z)=\frac{9}{1+z}\left(\frac{M}{1.5\times 10^{13}h^{-1}M_{\odot}}\right)^{-0.13}.
\end{equation}
While in \citet{clbf00}, the concentration is altered depending on the
history of each dark halo, we simply assume the above form of the
concentration.

A part of the hot gas cools and accretes to the disk of a galaxy until
subsequent collapse of dark halos containing this halo.  The amount of
the cooled gas is defined by the mass of the hot gas encircled by the
{\it cooling radius}, at which the time elapsed from the formation epoch
is equal to the cooling time-scale derived using the above density
distribution,
\begin{equation}
 t_{\rm cool}(r)=\frac{3}{2}\frac{\rho_{\rm hot}(r)}{\mu m_{\rm p}}
  \frac{k_{\rm B}T_{\rm vir}}{n_{\rm e}^{2}(r)\Lambda(T_{\rm vir}, Z_{\rm hot})},
\end{equation}
where $n_{\rm e}(r)$ are the electron density of hot gas at $r$, $Z_{\rm
hot}$ is the metallicity of hot gas, and $\Lambda$ is a
metallicity-dependent cooling function provided by \citet{sd93}.  For
simplicity, gas which cools in a time-step within a halo is added to
cold gas of the central galaxy of the halo, which is defined in \S\S
\ref{sec:merger}, at the beginning of the time-step.  Chemical
enrichment in hot gas is consistently solved with star formation and SN
feedback, as shown below.

In order to avoid the formation of unphysically large galaxies, the
cooling process is applied only to halos with $V_{\rm circ}\leq V_{\rm
cut}$.  Although \citet{clbf00} apparently succeeded in avoiding the
formation of {\it monster galaxies} without such a cooling cut-off by
introducing a history-dependent core radius, they have found that, in
the case of higher baryon density, $\Omega_{\rm b}\simeq 0.04$, too many
monster galaxies emerge \citep{bbflbc03}.  Here we take a simple
approach to avoid the formation of monster galaxies, which is stopping
cooling by hand.  This will be solved if we treat heating processes in
clusters of galaxies correctly, e.g. heating by AGN/QSOs.  Throughout
this paper, we set $V_{\rm cut}=210$ km s$^{-1}$.  This value might seem
somewhat small.  In addition to the higher baryon density, the reason
for the difference of this kind of cooling cut-off from ones used in
other recent SA models, in which $V_{\rm cut}\sim$ 350-500 km~s$^{-1}$
are used, is the treatment of reheated gas by SNe.  For example, {\it
superwinds} are adopted in \citet{baugh05} and \citet{nlbfc05} which
expel gas by SNe in addition to the usual SN feedback.  This frees us
from the artificial cooling cut-off.  However, the same model, but
without the superwind, must adopt a small value of circular velocity for
the cooling cut-off.  \citet{nlbfc05} have reported that a model without
the superwind requires $V_{\rm cut}=100 (1+z)^{3/4}$ km~s$^{-1}$, where
the redshift dependence originates by assuming heating by conductive
heat fluxes from outer envelopes of halos.  Thus, in the case of
$\Omega_{\rm b}=0.048$, under the assumption that all baryons are
contained in halos, such a small $V_{\rm cut}$ is required.  In future,
mechanisms of gas-heating and of gas-expulsion from halos should be
modeled.

The above assumption that all baryons are retained by dark halos might
be too simple.  Some have argued that gas known as the warm/hot
intergalactic medium (WHIM) may be outside halos \citep[e.g.][]{co99,
yoshikawa04, nicastro05}.  If the WHIM is the gas expelled from
galactic-scale halos caused by the superwind or similar mechanisms, the
baryon fraction in the halos is lowered.  Instead, the WHIM might be the
gas contained in smaller halos which are not resolved in hydrodynamical
simulations \citep[e.g.][]{insa04}.  Future observation of the WHIM will
provide a clue to the baryon fraction in galactic-scale halos, which
constrains the value of $V_{\rm cut}$.

\subsection{Star Formation and Supernova Feedback}\label{sec:sffb}
Stars in disks are formed from the cold gas.  The SF rate (SFR) $\psi$
is given by the cold gas mass $M_{\rm cold}$ and a SF time-scale
$\tau_{*}$ as $\psi=M_{\rm cold}/\tau_{*}$.  Although determining
$\tau_{*}$ in a purely theoretical way is quite difficult, recent SA
models suggest that $\tau_{*}$ should be dependent on characteristic
velocities of galaxies and independent of redshift to explain many
observations \citep{kh00, ntgy01, nytg02, eng03, ongy04, ny04, baugh05,
nlbfc05}.  In this paper, we adopt a simple prescription, similar to
that of \citet{cafnz94},
\begin{equation}
 \tau_{*}=\tau_{*}^{0}\left(\frac{V_{d}}{V_{\rm hot}}\right)^{\alpha_{*}},\label{eqn:taustar}
\end{equation}
where $\tau_{*}^{0}$ and $\alpha_{*}$ are free parameters, chosen to
match the ratio of the cold gas mass to the $B$-band luminosity of
spiral galaxies and $V_{d}$ is the disk rotation velocity defined in
\S\S \ref{sec:response}.  $V_{\rm hot}$ is a parameter related to the SN
feedback (see below).  In NY04, the star formation time-scale was
\begin{equation}
 \tau_{*}=\tau_{*}^{0}\left[1+\left(\frac{V_{d}}{V_{\rm
 hot}}\right)^{-\alpha_{\rm hot}}\right],
\end{equation}
where $\alpha_{\rm hot}$ is also a parameter related to the SN feedback
defined below.  Because this prescription has only one free parameter,
$\tau_{*}^{0}$, it is very helpful to search for suitable parameters.
Although this worked well in NY04, in fact, we found that this cannot
reproduce cold gas fractions in spiral galaxies in the case of weaker SN
feedback.

Massive stars explode as Type II SNe and heat up the surrounding cold
gas.  This SN feedback reheats the cold gas at a rate of $\dot{M}_{\rm
reheat}={M_{\rm cold}}/{\tau_{\rm reheat}}$, where the time-scale of
reheating is given by
\begin{equation}
\tau_{\rm reheat}=\frac{ \tau_{*}}{\beta(V_{d})},
\end{equation}
where
\begin{equation}
\beta(V_{d})\equiv\left(\frac{V_{d}}{V_{\rm hot}}
\right)^{-\alpha_{\rm hot}}.\label{eqn:vhot}
\label{eqn:beta}
\end{equation}
The free parameters $V_{\rm hot}$ and $\alpha_{\rm hot}$ are determined
by matching the local luminosity function of galaxies with observations.

Below, we shall use two parameter sets.  One of them has a strong SN
feedback efficiency with $\alpha_{\rm hot}=4$.  This model is hereafter
called the ``strong feedback'' (SFB) model.  In this case, because of
the strong SN feedback, we can take $\alpha_{*}=-\alpha_{\rm hot}$,
which makes a star formation time-scale similar to NY04.  It is not the
case for a weak SN feedback.  In the other model, the SN feedback is
weak, $\alpha_{\rm hot}=2$.  This is called the ``weak feedback'' (WFB)
model.

With the above equations and parameters, we obtain the masses of hot
gas, cold gas, and disk stars as a function of time or redshift.  

Associated with star formation and SN feedback, the chemical enrichment
is also taken into account by extending \citet{maeder92}.  For
simplicity, instantaneous recycling is assumed for SNe II, and any
contribution from SNe Ia is neglected.  The recent progress on the
chemical enrichment with SNe Ia in the framework of SA models is found
in \citet{no04} and \citet{nlbfc05, nlobfc05}.

To sum up the above, the basic equations are
\begin{eqnarray}
 \dot{M}_{*}&=&\alpha\psi,\\
 \dot{M}_{\rm cold}&=&-(\alpha+\beta)\psi,\\
 \dot{M}_{\rm hot}&=&\beta\psi,\\
 (M_{\rm cold}Z_{\rm cold})\dot{}&=&[p-(\alpha+\beta)Z_{\rm cold}]\psi,\label{eqn:p}\\
 (M_{\rm hot}Z_{\rm hot})\dot{}&=&\beta Z_{\rm cold}\psi,
\end{eqnarray}
where the dot ($\dot{}$~) indicates the time derivative, $M_{*}$ and
$M_{\rm hot}$ are the masses of stars and hot gas, respectively,
$\psi={M_{\rm cold}}/{\tau_{*}}$ is the SFR, $\alpha$ is a locked-up
mass fraction which is set to be 0.75, consistent with a stellar
evolution model we use, $Z_{\rm cold}$ and $Z_{\rm hot}$ are the
metallicities of cold and hot gases, respectively, and $p$ is the
chemical yield which is set to be twice the solar metallicity in order
to be consistent with the observed metallicity distribution of
elliptical galaxies.  We can solve these equations analytically as
\begin{eqnarray}
 \Delta M_{\rm cold}(t)&=&M_{\rm cold}^{0}\left\{1-\exp\left[-(\alpha+\beta)\frac{t}{\tau_{*}}\right]\right\},\\
 \Delta M_{*}(t)&=&\frac{\alpha}{\alpha+\beta}\Delta M_{\rm cold}(t),\\
 \Delta M_{\rm hot}(t)&=&\frac{\beta}{\alpha+\beta}\Delta M_{\rm cold}(t),\\
 Z_{\rm cold}(t)&=&Z_{\rm cold}^{0}+p\frac{t}{\tau_{*}},\\
 Z_{\rm hot}(t)&=&\left[M_{\rm hot}^{0}Z_{\rm
 hot}^{0}+\frac{\beta}{\alpha+\beta}\left\{\frac{}{}\right.\right.\nonumber\\
 &&\left(\frac{p}{\alpha+\beta}+Z_{\rm
 cold}(t)\right)\Delta M_{\rm cold}(t) \nonumber\\
 &&-(Z_{\rm cold}(t)-Z_{\rm
 cold}^{0})M_{\rm cold}^{0}{\left.\left.\frac{}{}\right\}\right]}/M_{\rm hot}(t),
\end{eqnarray}
where variables with $\Delta$ indicate increments or decrements which
are defined as positive and the superscript 0 stands for initial values
at the beginning of the time-step, at which time $t$ in the solutions is
set at zero.  For reference, we describe a mass-weighted mean stellar
metallicity because it should be helpful to see the role of SN feedback
on the chemical enrichment.  The mass-weighted metallicity is given by
\begin{eqnarray}
 \langle Z_{*}(t)\rangle&=&\frac{\int_{0}^{t}\dot{M}_{*}Z_{\rm cold}(t)dt}{\int_{0}^{t}\dot{M}_{*}dt}\nonumber\\
 &=&Z_{\rm cold}^{0}+\frac{p}{\alpha+\beta}\frac{1-e^{-u}-ue^{-u}}{1-e^{-u}},
\end{eqnarray}
where $u\equiv (\alpha+\beta)t/\tau_{*}$.  This means that when the SN
feedback is strong, the increase in mean stellar metallicity is small.
The SN feedback clearly plays an important role in chemical enrichment.

\subsection{Mergers of galaxies and formation of spheroids}\label{sec:merger}
When two or more progenitor halos have merged, the newly formed larger
halo should contain at least two or more galaxies which had originally
resided in the individual progenitor halos.  By definition, we identify
the central galaxy in the new common halo with the central galaxy
contained in the most massive of the progenitor halos.  Other galaxies
are regarded as satellite galaxies.  These satellites merge either by
dynamical friction or random collision.  The time-scale of merging by
dynamical friction is given by $\tau_{\rm mrg}=f_{\rm mrg}\tau_{\rm
fric}$, where $\tau_{\rm fric}$ is given by \citet{bt87},
\begin{equation}
 \tau_{\rm fric}=\frac{1.17}{\ln\Lambda_{\rm C}}\frac{R_{h}^{2}V_{\rm circ}}{GM_{\rm sub}},
\end{equation}
where $R_{h}$ is the radius of the new common halo, $M_{\rm sub}$ is the
mass of the tidally truncated subhalo, and $\Lambda_{\rm C}$ is the
Coulomb logarithm.  The parameter $f_{\rm mrg}$ is set to 0.8 in this
paper by matching the bright end of luminosity functions.  In order to
judge whether or not a satellite merges with a central galaxy, we
compute the time elapsed for each satellite galaxy since the galaxy
becomes a satellite.  This elapsed time is reset to zero if the mass of
the host halo doubles.  When the elapsed time exceeds $\tau_{\rm mrg}$,
which is estimated from masses of the host halo and a tidally-stripped
subhalo harboring the satellite, the satellite galaxy is accreted to the
central galaxy.

Satellite galaxies sometimes merge in a time-scale of random collision.
The merger rate, $k$, can be written using physical parameters of
galaxies and their host halo,
\begin{equation}
 k\sim n\sigma V_{\rm circ},
\end{equation}
where $n$ is the number density of galaxies, and $\sigma$ the cross
section.  Since the cross section is estimated from gravitational
focusing, we can write $\sigma\propto (Gm_{g}/V_{\rm circ}^{2})^{2}$,
where $m_{g}$ is the mass of each galaxy.  Using the virial relation,
$m_{g}\simeq r_{g}v_{g}^{2}/G$, where $r_{g}$ is the size of galaxies
and $v_{g}$ a characteristic velocity of galaxies, the merger rate is
\begin{equation}
 k\propto \frac{N}{R_{h}^{3}} \frac{r_{g}^{2}v_{g}^{4}}{V_{\rm circ}^{3}},
\end{equation}
where $N$ is the number of galaxies in the halo.  \citet{mh97} derived a
merger rate in a situation in which the mass of all galaxies was the
same,
\begin{eqnarray}
 k_{\rm MH}&=&\frac{N}{500}\left(\frac{1\rm Mpc}{R_{h}}\right)^{3}
  \left(\frac{r_{g}}{0.1\rm Mpc}\right)^{2}\nonumber\\
  &&\times\left(\frac{\sigma_{g}}{100\rm km~s^{-1}}\right)^{4}
  \left(\frac{300\rm km~s^{-1}}{\sigma_{h}}\right)^{3} \mbox{Gyr}^{-1},
\end{eqnarray}
where $\sigma_{g}$ and $\sigma_{h}$ are the one-dimensional velocity
dispersions.  Note that the {\it total} merger rate is $R_{\rm
tot}=Nk_{\rm MH}$.  Unless explicitly mentioned, we use this Makino-Hut
merger rate for satellite-satellite mergers, as in \citet{sp99}.  After
that, \citet{m00} has proposed relaxing the assumption of equal-mass
mergers in the following way: Consider a merger of galaxies with masses
$m$ and $\lambda m$ and assume $m\sim r_{g}^{3}$.  Because $k$ is
considered to scale as $\langle r_{g}\rangle^{2}\langle
v_{g}^{2}\rangle^{2}$, the rate should be
\begin{equation}
 k(m,\lambda m)=k\left(\frac{1+\lambda^{1/3}}{2}\right)^{2}
  \left(\frac{1+\lambda^{2/3}}{2}\right)^{2}.\label{eqn:mamon}
\end{equation}
To introduce this into our model, the merger rate of a galaxy of mass
$m$ with galaxies of masses $\lambda_{i}m$, where $i$ stands for other
satellite galaxies in the same halo, is
\begin{equation}
 \langle k(m)\rangle=\sum_{i}^{N-1}\frac{k(m,\lambda_{i}m)}{N}.\label{eqn:mamon2}
\end{equation}
Later we shall see how this effect of non-equal mass mergers affects our
results in Figure \ref{fig:lf4}.  In most of our results, however, we do
not take into account this effect because it is almost negligible, at
least at present, and because this requires quite a long time to
compute.  Below, we take $k_{\rm MH}$ as a fiducial merger rate, $k$, for
equal-mass mergers, and use quantities of subhalos for the mass, radius
and velocity dispersion of satellite galaxies.  Here we define a
collision time-scale, $\tau_{\rm coll}=1/\langle k\rangle$.  With a
probability $\Delta t/\tau_{\rm coll}$, where $\Delta t$ is the
time-step corresponding to the redshift interval $\Delta z$ of merger
trees of dark halos, a satellite galaxy merges with another satellite
picked out randomly.

Consider the case when two galaxies of masses $m_1$ and $m_2 (>m_1)$
merge.  If the mass ratio $f=m_1/m_2$ is larger than a certain critical
value of $f_{\rm bulge}$, we assume that a starburst occurs and that all
of the cold gas turns into stars and hot gas according to the same SN
feedback law, eq.(\ref{eqn:beta}), and all of the stars populate the
bulge of a new galaxy.  Note that when applying the SN feedback law,
$V_{d}$ is replaced by $V_{b}$ defined in \S\S \ref{sec:response}, which
is the velocity dispersion of the new bulge.  The time-scale of the
starburst is given by the dynamical time-scale, $\tau_{*}=r_{e}/V_{b}$,
where $r_{e}$ is the effective radius of the bulge, also defined in \S\S
\ref{sec:response}.  On the other hand, if $f<f_{\rm bulge}$, no
starburst occurs, and a smaller galaxy is simply absorbed into the bulge
of a larger galaxy.  Note that in NY04 the smaller galaxy was absorbed
into the disk of the larger galaxy.  We have changed it to resemble the
observed distribution of bulge-to-disk ratios of galaxies.  Throughout
this paper we use $f_{\rm bulge}=0.3$, which gives a consistent
morphological fraction in $I_{814}$-band galaxy number counts [see,
e.g., Figure 13 in \citet{nytg02}].

\subsection{Size of Galaxies and Dynamical Response to Starburst-induced Gas Removal}\label{sec:response}

\begin{figure*}
\plotone{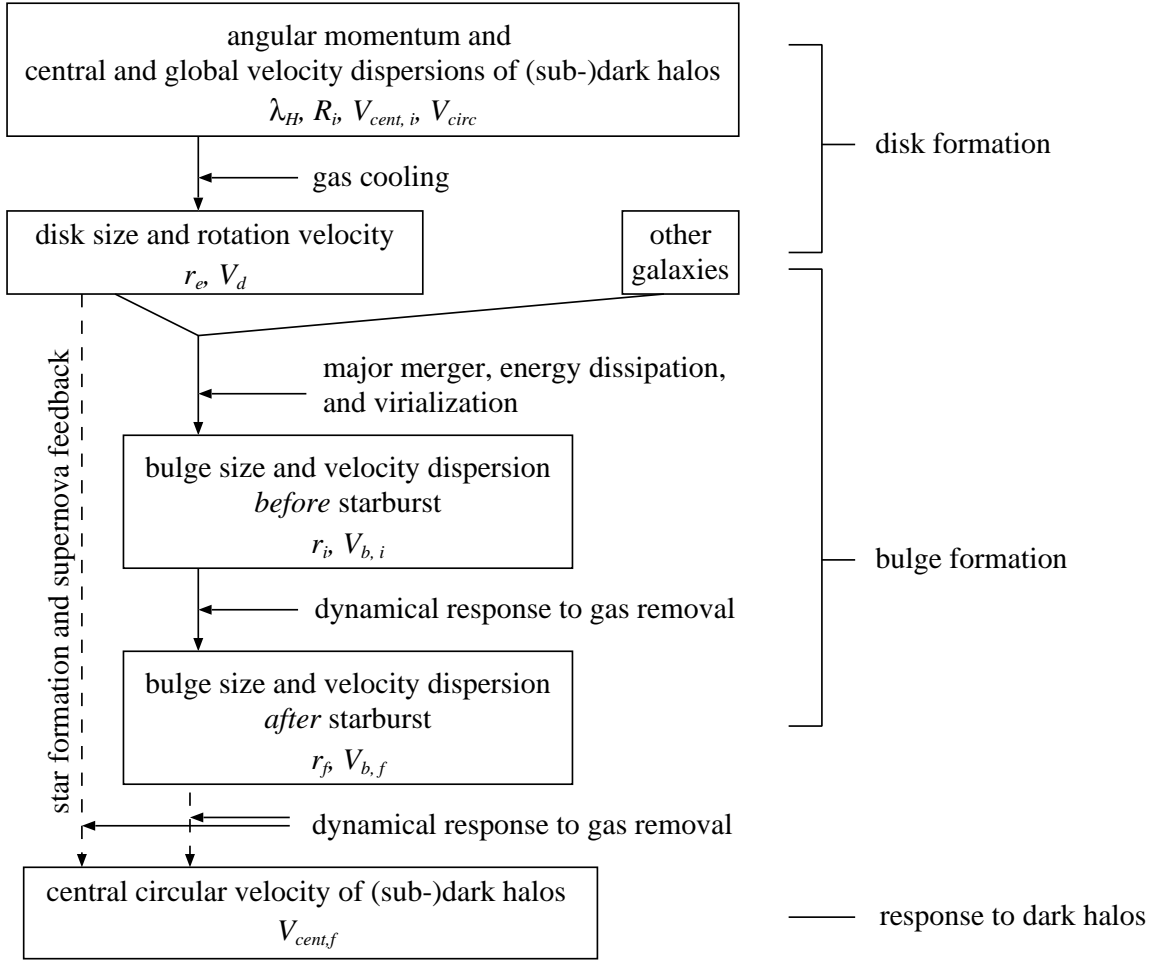}

\caption{Diagram showing the flow to estimate sizes and velocities of
 dark halos, disks, and bulges.}

\label{fig:sizescheme}
\end{figure*}

Here we show a full description to estimate the size of galaxies, the
rotation velocity of disks, and the velocity dispersion of bulges.  The
procedure is almost the same as that given by NY04 and shown
schematically in Figure \ref{fig:sizescheme}.  A new parameter providing
a fraction of energy dissipation during the major merger is introduced.
It should be noted that the inclusion of the dynamical response to gas
removal induced by starbursts provides much better agreement with
observed kinematical and structural parameters, particularly for dwarf
galaxies, as shown in NY04.

Dark halos have been considered to acquire angular momenta during the
linear regime of their collapse through tidal torques made by the
initial density fluctuation field \citep{p69, w84, ct96a, ct96b, ng98}.
Assuming that the hot gas component that cools and contracts within a
dark halo has the same specific angular momentum as the dark halo has,
we can estimate a radius at which the gas is supported by rotation with
the conservation of its specific angular momentum.  In this way, we
determine the sizes of spiral galaxies.

In this model, angular momenta of dark halos are given by the following
procedure, independent of the $N$-body results: This is not only for
simplicity, but also because it is challenging at this stage to deal
with the change in direction of the angular momentum, which is neglected
here as usually assumed.  The distribution of the dimensionless spin
parameter $\lambda_{\rm H}$, which is defined by $\lambda_{\rm H}\equiv
L|E|^{1/2}/GM^{5/2}$ where $L$ is the angular momentum and $E$ is the
binding energy, is well approximated by a log-normal distribution
\citep{mmw98},
\begin{equation}
 p(\lambda_{\rm H})d\lambda_{\rm H}=
\frac{1}{\sqrt{2\pi}\sigma_{\lambda}}
\exp\left[-\frac{(\ln\lambda_{\rm H}-\ln\bar{\lambda})^2}
{2\sigma_{\lambda}^{2}}\right] d\ln\lambda_{\rm H},\label{eqn:lognormal}
\end{equation}
where $\bar{\lambda}$ is the mean value of the spin parameter and
$\sigma_{\lambda}$ is its logarithmic variance.  We adopt
$\bar{\lambda}=0.05$ and $\sigma_{\lambda}=0.5$ according to
\citet{mmw98}.  When the angular momentum of the hot gas is conserved,
the effective radius $r_{e}$ of a resultant cold gas disk is related to
the initial radius $R_{i}$ of the hot gas sphere via
$r_{e}=(1.68/\sqrt{2})\lambda_{\rm H}R_{i}$ \citep{f79, fe80, f83}.  The
initial radius $R_{i}$ is set to be the smaller of the virial radius of
the host halo and the cooling radius.  A galaxy disk grows due to
cooling and accretion of hot gas from the more distant envelope of its
host halo.  In our model, the disk size of each central galaxy is
estimated at each time-step by the above method, but it is not always
renewed.  Only when both mass and estimated size of each disk are larger
than those in the previous time-step, the size is renewed.  At this
time, the disk rotation velocity $V_{d}$ is set to be the circular
velocity of its host dark halo.

Size estimation of high-redshift spiral galaxies, however, carries
uncertainties because of the large dispersion in their observed size
distribution.  For example, \citet{s99} suggests only mild evolution of
disk size against the redshift, taking into account the selection
effects arising from the detection threshold of surface brightness,
although the above simple model predicts disk size proportional to
virial radius $R_{\rm vir}$ of host dark halos evolving as $R_{\rm
vir}\propto 1/(1+z)$ for a fixed mass.  Allowing for the possibility
that the conservation of the angular momentum is not complete and/or for
other sources such as the SN feedback to stop the contraction of gas
disks, we generalize this size estimation by introducing a simple
redshift dependence,
\begin{equation}
 r_{e}=\frac{1.68}{\sqrt{2}}\lambda_{\rm H}R_{i}(1+z)^{\rho},
\label{eqn:sizerho}
\end{equation}
where $\rho$ is a free parameter.  We simply use $\rho=1$ as a reference
value in this paper in order to be consistent with the observational
galaxy number counts.  The changing of the value of $\rho$ affects the
selection effects due to the cosmological dimming of surface brightness
and the dust extinction, because the dust column density also changes
with galaxy size.

In the case of the formation of elliptical galaxies driven by major
mergers of galaxies, the size and velocity dispersion are estimated,
explicitly taking into account the dynamical response to
starburst-induced gas removal.  The outline is as follows: At first, two
or more galaxies merge.  During the merger, a fraction $f_{\rm diss}$ of
the total energy of baryonic matter is assumed to be dissipated and a
spheroidal merger remnant is formed and reaches the virial equilibrium
immediately.  Subsequently, the starburst occurs and a fraction of gas is
gradually expelled from the spheroidal system according to the SN
feedback law.  Because the gravitational potential varies during the gas
removal, the system expands and its velocity dispersion is lowered.  The
detail is shown below.

Sizes of early-type galaxies are primarily determined by the virial
radius of the baryonic component.  When a major merger of galaxies
occurs, assuming some fraction of the energy loss to the total energy
and no rotation of the merger remnant, we estimate the velocity
dispersion of the merged system.  Now we assign the subscript 0 to the
merged galaxy, and subscripts 1 and 2 to the central and satellite
galaxies, respectively, in the case of central-satellite merger, or to
larger and smaller galaxies, respectively, in the case of
satellite-satellite merger.  Using the virial theorem, the total energy
for each galaxy is
\begin{equation}
 E_{i}=-\frac{1}{2}[M_{b}V_{b}^{2}+(M_{d}+M_{\rm cold})V_{d}^{2}],
\end{equation}
where $M_{b}$ and $M_{d}$ are the masses of bulge and disk stars,
respectively, and $V_{b}$ and $V_{d}$ are the velocity dispersion of
bulge and the rotation velocity of disk, respectively.  Assuming the
virial equilibrium, the binding energy $E_{b}$ between the progenitors
just before the merger is given by
\begin{equation}
 E_{b}=-\frac{E_{1}E_{2}}{(M_{2}/M_{1})E_{1}+(M_{1}/M_{2})E_{2}}.
\end{equation}
Then we obtain
\begin{equation}
 f_{\rm diss}(E_{1}+E_{2}+E_{b})=E_{0},\label{eqn:fdiss}
\end{equation}
where $f_{\rm diss} (\geq 1)$ is a parameter which describes how large a
fraction is dissipated from the system.  Just after the merger, there
remains only the bulge component consisting of cold gas and stars in the
merger remnant, whose velocity dispersion is directly estimated from the
above equation.  Then the size of the system just after the merger is
defined by
\begin{equation}
 r_{i}=\frac{GM_{i}}{2V_{b}^{2}},\label{eqn:sizeE}
\end{equation}
where $M_{i}=M_{*}+M_{\rm cold}$ is the total baryonic mass of the
merged system.  In reality, the rate of energy dissipation depends on
complicated physical processes such as the escape of high velocity stars
and the viscosity and friction due to gas.  Throughout this paper,
however, we adopt $f_{\rm diss}=1.3$ for simplicity, which is chosen so
that the size and velocity dispersion of ellipticals agrees with
observations.  Apart from $f_{\rm diss}$, this way of estimating $V_{b}$
and $r_{i}$ is similar to that adopted in other SA models such as that
by \citet{clbf00}, in which they formulate the conservation of energy in
terms of the size.

Next, the cold gas turns into stars and hot gas.  Since part of the gas
is expelled from galaxies to halos due to the SN feedback, the mass of
the system changes.  Assuming the density distributions of baryonic and
dark matters, the dynamical response on the structural parameters to the
mass loss can be estimated.  This process was taken into account for the
first time in the Mitaka model (NY04) in the framework of SA modeling.
In this paper we adopt the Jaffe model \citep{jaffe} for baryonic matter
and the static singular isothermal sphere for dark matter, and assume
slow (adiabatic) gas removal compared with the dynamical time-scale of
the system.  Note that \citet{ny03} also show the dynamical response for
different density distributions of baryonic and dark matter and for
rapid gas removal cases.  Defining by ${\cal M}, R, Y$ and $U$ the
ratios of mass, size, density and velocity dispersion at a final state
relative to those at an initial state, the response under the above
assumption is approximately given by
\begin{eqnarray}
R&\equiv&\frac{r_{f}}{r_{i}}=\frac{1+D/2}{{\cal M}+D/2},\label{eqn:sizeE2}\\
U&\equiv&\frac{V_{b,f}}{V_{b,i}}=\sqrt{\frac{YR^{2}+Df(z_{f})/2}{1+Df(z_{i})/2}},
\end{eqnarray}
where ${\cal M}=YR^{3}$, $D=1/y_{i}z_{i}^{2}$, $y$ and $z$ are the
ratios of density and size of baryonic matter to those of dark matter
and, $f(z)$ is 
\begin{equation}
 f(z)=\frac{\ln(1+z)}{z}+\ln\left(1+\frac{1}{z}\right).
\end{equation}
The subscripts $i$ and $f$ stand for the initial and final states in the
mass loss process.   The contribution of dark matter is estimated
from the central circular velocity of halos, $V_{\rm cent}$, which is
defined below.  According to numerical simulations of starbursts
performed by \citet{mytn} and \citet{myn}, the time-scale of gas removal
might be comparable to the dynamical time-scale for dwarf ellipticals.
If the gas-removal is almost instantaneous, the effect of the dynamical
response becomes stronger, as shown by \citet{ny03}.  Therefore, the
assumption of the slow gas removal is conservative.  In reality, some
galaxies might become gravitationally unbound due to the rapid
gas-removal.  Stars of such galaxies should contribute to halo stars or
intracluster stars.  

Finally the back reaction to the gas removal by the SN feedback on dark
matter distribution is computed.  It is natural to consider that dark
matter distribution in the central region of dark halos is also affected
by the dynamical response to gas removal, because baryons usually
condense with density comparable to that of dark matter.  To take into
account this process, we define a central circular velocity of dark halo
$V_{\rm cent}$.  When a dark halo collapses without any progenitors,
$V_{\rm cent}$ is set to $V_{\rm circ}$.  After that, although the mass
of the dark halo grows by subsequent accretion and/or mergers, $V_{\rm
cent}$ remains constant or decreases with the dynamical response.  When
the mass is doubled, $V_{\rm cent}$ is set to $V_{\rm circ}$ again.  The
dynamical response to mass loss from a central galaxy of a dark halo by
SN feedback lowers $V_{\rm cent}$ of the dark halo as follows:
\begin{equation}
 \frac{V_{{\rm cent},f}}{V_{{\rm cent},i}}=
  \frac{M_{f}/2+M_{d}(r_{i}/r_{d})}{M_{i}/2+M_{d}(r_{i}/r_{d})}.
\end{equation}
The change of $V_{\rm cent}$ in each time-step is only a few per cent.
Under these conditions, the approximation of static gravitational
potential of dark matter is valid during starbursts.   This also
applies to subhalos.

Once a dark halo falls into its host dark halo, it is treated as a
subhalo.  Because we assume that subhalos do not grow in mass, the
central circular velocity of the subhalos monotonically decreases.
Thus, this affects the dynamical response later when mergers between
satellite galaxies occur.  We assume that the resultant density
distribution remains isothermal with $V_{\rm cent}$ at least within the
galaxy size.

The details of the dynamical response are shown in \citet{ny03, ny04}.
The effect of the dynamical response is the most prominent for dwarf
galaxies of low circular velocity because of the substantial removal of
gas due to strong SN feedback \citep{ya87, ny04}.  If the dynamical
response had not been taken into account, velocity dispersions of dwarf
ellipticals would have been much larger than those of observations,
determined only by circular velocities of small dark halos in which
dwarf ellipticals resided.  For giant ellipticals, on the other hand,
the effect of the dynamical response is negligible because only a small
fraction of gas can be expelled due to weak SN feedback.  We refer the
reader to NY04 for detailed discussion of the dynamical response.

\subsection{Photometric Properties and Morphological Identification}\label{sec:photo}
The above processes are repeated until the output redshift and then the
SF history of each galaxy is obtained.  For the purpose of comparison
with observations, we use a stellar population synthesis approach, from
which the luminosities and colors of model galaxies are calculated.
Given the SFR as a function of time or redshift, the absolute luminosity
and colors of individual galaxies are calculated using a population
synthesis code by \citet{ka97}.  The stellar metallicity grids in the
code cover a range from $Z_{*}=$0.0001 to 0.05. Note that we now define
the metallicity as the mass fraction of metals.  The initial stellar
mass function (IMF) that we adopt is the power-law IMF of Salpeter form
\citep{s55}, with lower and upper mass limits of $0.1M_{\odot}$ and
$60M_{\odot}$, respectively.  In some SA models, it has been assumed
that there is a substantial fraction of invisible stars such as brown
dwarfs which have masses smaller than 0.1$M_{\odot}$.  In this paper,
however, we do not assume any contributions of such invisible stars to
the stellar mass according to NY04.

The optical depth of internal dust is consistently estimated by our SA
model.  We make the usual assumption that the abundance of dust is
proportional to the metallicity of cold gas, and the optical depth is
proportional to the column density of metals.  Then the optical depth
$\tau$ is given by
\begin{equation}
 \tau\propto\frac{M_{\rm cold}Z_{\rm cold}}{r_{e}^{2}}(1+z)^{-\gamma},
\label{eqn:dust}
\end{equation}
where $r_{e}$ is the effective radius of galaxies and $\gamma$ is a free
parameter which should be chosen to predict high redshift galaxies being
consistent with observations.  There are large uncertainties in
estimating the proportionality constant, but we adopt about half the
value used by \citet{clbf00} according to NY04, because it predicts too
strong an extinction to reproduce galaxy number counts.  Wavelength
dependence of optical depth is assumed to be the same as the Galactic
extinction curve given by \citet{seaton79}.  Dust distribution is simply
assumed to obey the slab dust model \citep{ddp89} for disks, according
to our previous papers.

In this paper, we introduce dust extinction during starbursts.  We
randomly assign the merger epoch within a time-step under consideration,
according to the time-scale of mergers.  The SF time-scale of the
bursting galaxy is assumed to be the dynamical time-scale,
$r_{e}/V_{b}$.  This time-scale enables us to compute the amount of
metals in cold gas at the output redshift or the end of the time-step.
Assuming the screen dust model, the optical depth is estimated.  This
procedure is important to obtain consistency with observational colors
of high redshift galaxies.  If we do not include this, some galaxies
right after major mergers become too blue to be consistent with
observations.  This is discussed in \S\S\ref{sec:color}.

The origin of $\gamma$ is still unknown.  We have introduced this
parameter just as a phenomenological one.  In the WFB model, we set
$\gamma=0$, but in the SFB model $\gamma=1$ to decrease the optical
depth at high redshift, otherwise too few galaxies were formed to
reconcile with observed galaxy number counts.  One of the reasons this
is required is that the dust-to-gas ratio may evolve with redshift.  At
high redshift, the ratio might be lower than at low redshift.  Another
possible reason is that the chemical yield we adopt is nearly equal to
twice the solar value, which might be too large.  This large value is
required to be consistent with observed colors of cluster elliptical
galaxies, which have been considered to have shallower IMFs of stars
than disk stars, so that the chemical yield may be larger than the solar
\citep{ay86, ay87}.  If this is true, the value of the chemical yield
must be different depending on the environment, that is, the yield for
disk stars should be approximately solar, but larger than solar for
spheroidal components.  Because $\gamma$ is required to decrease the
optical depth of dust in disks, by adopting a chemical yield similar to
the solar value, we will be able to set $\gamma=0$.  As for recent
progress in investigating the chemical enrichment using SA models, we
refer the reader to \citet{no04}, \citet{baugh05} and \citet{nlbfc05}.
In this paper, we do not enter into any details and maintain $\gamma$ as
a phenomenological parameter to reproduce observed high redshift
galaxies.

\begin{figure}
\plotone{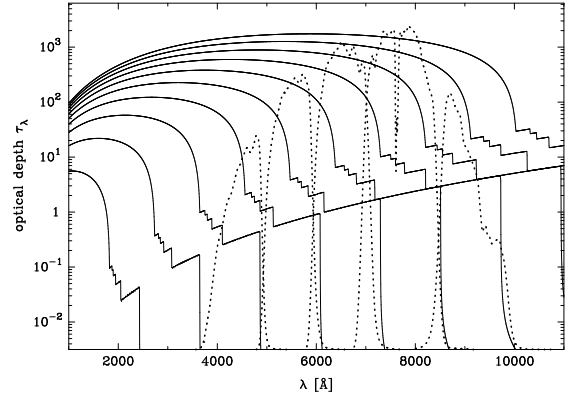}

\caption{Optical depth as a function of observed wavelength for various
source redshifts, $z_{s}=1, 2, ... 10$ from the left to the right.
Response functions of $BVRi'z'$ bands for the Suprime-Cam on the Subaru
telescope are overlaid by the dotted lines from the left to the right.
Note that the vertical axis for these is in linear scale.  }

\label{fig:tau}
\end{figure}

Light emitted from high redshift galaxies is also absorbed by
intergalactic \ion{H}{1} clouds.  We include this effect in our SA
model, according to \citet{yp94} \citep[see also][]{madau95, madau96}.
The optical depth for a source at redshift $z_{s}$ at a wavelength
$\lambda$ is written as
\begin{equation}
 \tau_{\lambda}(z_{s})=\int_{0}^{z_{s}}\int_{0}^{\infty}dW\frac{\partial^{2}{\cal
 N}}{\partial z\partial W}\left\{1-e^{-s[(1+z)\nu]N(W)}\right\},
\end{equation}
where $\nu=c/\lambda$, ${\partial^{2}{\cal N}}/{\partial z\partial W}$
is the number of the Ly$\alpha$ lines per unit interval redshift per
rest-frame Ly$\alpha$ equivalent width $W$, $N(W)$ is the hydrogen
column density which is related to $W$ through the curve of growth for
the Ly$\alpha$ absorption line, and $s(\nu)$ is the sum of absorption
cross sections for the lines and continuum of the hydrogen atom
dependent on the Doppler $b$-parameter, which is assumed to be $b=20$ km
s$^{-1}$.  The cross sections for the lines and the continuum are
expressed as
\begin{eqnarray}
\lefteqn{s_{nn'}(\nu)=\frac{\pi e^{2}}{m_{e}c}f_{nn'}}\nonumber\\
 &&\times\int_{-\infty}^{\infty}\frac{(\Gamma_{n'}/4\pi^{2})e^{-v^{2}/b^{2}}dv}
 {\sqrt{\pi}b\{[\nu_{nn'}+(v/c)\nu_{nn'}-\nu]^{2}+(\Gamma_{n'}/4\pi^{2})\}},
\end{eqnarray}
and
\begin{eqnarray}
 s_{n}(\nu)&=&\frac{64\pi^{4}m_{e}e^{10}g_{n}(\nu)}{3\sqrt{3}ch^{6}n^{5}\nu^{3}}\nonumber\\
  &\times&\int_{-\infty}^{\infty}\frac{\{1-\theta[\nu_{n}+(v/c)\nu_{n}-\nu]\}e^{-v^{2}/b^{2}}dv}
  {\sqrt{\pi}b},
\end{eqnarray}
where $\theta(x)$ is the Heaviside step function, $\Gamma_{n'}$ is the
effective damping constant, $g_{n}(\nu)$ is the Gaunt factor, and the
other notations have their usual meanings.  We consider Lyman continuum
absorption ($n=1$) and the first five lines of Lyman series,
Ly$\alpha$, ... Ly$\epsilon$ ($n=1$ and $n'=2, ... 6$).  The line
distribution can be modeled as
\begin{equation}
 \frac{\partial^{2}{\cal N}}{\partial z\partial W}=10.7(1+z)^{2.37}\frac{e^{-W/W^{*}}}{W^{*}},
\end{equation}
where $W^{*}=0.36$\AA, according to the Model I in \citet{yp94}.
Combined with response functions of the filters we use, we obtain the
optical depths for individual pass-bands.  We simply estimate an optical
depth for an $X$-band, $\tau_{X}$, as
\begin{equation}
 \exp(-\tau_{X})=\frac{\int d\lambda R_{X,\lambda}e^{-\tau_{\lambda}}}
  {\int d\lambda R_{X,\lambda}},
\end{equation}
where $R_{X,\lambda}$ is a response function to the $X$-band filter.
The difference in magnitudes is given by $\Delta m=2.5\tau_{X}\log e$.
Figure \ref{fig:tau} shows the optical depths at $z_{s}=1, ... 10$,
indicated by the solid lines, and response functions of $BVRi'z'$ of the
{\it Subaru} telescope by the dotted lines, which are used in
\S\ref{sec:highz}.

\begin{deluxetable*}{lcclll}
\tabletypesize{\scriptsize}
\tablecaption{Astrophysical Parameters\label{tab:astro}}  
\tablewidth{0pt}
\tablehead{
\colhead{Parameter} & \colhead{SFB} & \colhead{WFB} & & \colhead{Annotation} & \colhead{Observation}}
\startdata
\hspace{-2ex}
$\begin{array}{ll}
~\alpha_{\rm hot}\\
~V_{\rm hot} ~(\rm{km~s}^{-1})
\end{array}$
&
\hspace{-2ex}
$\begin{array}{cc}
4\\
140
\end{array}$
&
\hspace{-2ex}
$\begin{array}{cc}
2\\
140
\end{array}$
&
$\left. \begin{array}{ll}
{}\\
{}
\end{array}\right\}$
 & supernova feedback-related (\S\S2.3) & luminosity functions (Figure \ref{fig:lf1}) \\
$V_{\rm cut}$ (km~s$^{-1}$) & 210 & 210
& & cooling cut-off (\S\S2.3) & luminosity functions (Figure \ref{fig:lf1})\\
\hspace{-2ex}
$\begin{array}{ll}
~\alpha_{*}\\
~\tau_{*}^{0} \mbox{(Gyr)}
\end{array}$
&
\hspace{-2ex}
$\begin{array}{cc}
-4\\
2.4
\end{array}$
&
\hspace{-2ex}
$\begin{array}{cc}
-2.3\\
2
\end{array}$
&
$\left. \begin{array}{ll}
{}\\
{}
\end{array}\right\}$
 & star formation-related (\S\S \ref{sec:sffb}) & cold gas mass (Figure \ref{fig:cold})\\
$p$ ($Z_{\odot}$) & 2 & 2 
& & heavy-element yield (\S\S \ref{sec:sffb}) & metallicity distribution
 (not shown)\\
$f_{\rm bulge}$ & 0.3 & 0.3
& & major/minor merger criterion (\S\S \ref{sec:merger}) & morphological
counts (not shown)\\
$f_{\rm mrg}$ & 0.8 & 0.8 
& & coefficient of dynamical friction timescale
 (\S\S \ref{sec:merger}) & luminosity functions (Figure \ref{fig:lf1}) \\
$f_{\rm diss}$ & 1.3 & 1.3 
& & energy-loss fraction (\S\S \ref{sec:response}) & spheroidal size (Figure \ref{fig:rad}) and
 velocity dispersion (Figure \ref{fig:fj})\\
\hspace{-2ex}
$\begin{array}{ll}
~\bar{\lambda}\\
~\sigma_{\lambda}
\end{array}$
&
\hspace{-2ex}
$\begin{array}{cc}
0.03\\
0.5
\end{array}$
&
\hspace{-2ex}
$\begin{array}{cc}
0.03\\
0.5
\end{array}$
&
$\left. \begin{array}{ll}
{}\\
{}
\end{array}\right\}$
 &
spin parameter distribution (\S\S \ref{sec:response}) & disk size (Figure \ref{fig:r})\\
$\rho$ & 1 & 1
& & redshift dependence of disk size (\S\S \ref{sec:response}) & faint galaxy
 number counts (Figures \ref{fig:sdf1}, \ref{fig:sdf2} and \ref{fig:angsize})\\
$\gamma$ & 1 & 0 
& & redshift dependence of dust optical depth (\S\S \ref{sec:photo}) &
 faint galaxy number counts (Figures \ref{fig:sdf1}, \ref{fig:sdf2} and \ref{fig:angsize})
\enddata

\tablecomments{ $\alpha_{\rm hot}$ and $V_{\rm hot}$ are the SN
feedback-related parameters (eq. \ref{eqn:vhot}).  $\alpha_{*}$ and
$\tau_{*}^{0}$ are the SFR-related parameters (eq. \ref{eqn:taustar}).
$f_{\rm bulge}$ is a critical mass-ratio of merging galaxies to
distinguish major and minor mergers (\S\S \ref{sec:merger}).  $f_{\rm
mrg}$ is a parameter to fine-tune the time-scale of mergers due to
dynamical friction (\S\S \ref{sec:merger}).  $p$ is the chemical yield
as defined by \citet{maeder92} (eq.  \ref{eqn:p}).  $\bar{\lambda}$ and
$\sigma_{\lambda}$ characterize the distribution function of the
dimensionless spin parameter, $\lambda_{\rm H}$, with a shape of
log-normal type (eq.  \ref{eqn:lognormal}).  Their values are given by
\citet{mmw98}.  $\rho$ is a parameter giving a redshift dependence of
the disk size (eq.  \ref{eqn:sizerho}).  $f_{\rm diss}$ represents the
energy-loss to the total energy during the major merger
(eq. \ref{eqn:fdiss}).  $\gamma$ is a parameter giving a redshift
dependence of dust optical depth (eq. \ref{eqn:dust}).  The last column
denotes observational quantities which are used to set the values of the
parameters of the model.  }
\end{deluxetable*}

We classify galaxies into different morphological types according to 
the $B$-band bulge-to-disk luminosity ratio $B/D$.  In this paper, 
following \citet{sdv86}, galaxies with $B/D\geq 1.52$, $0.68\leq 
B/D<1.52$, and $B/D<0.68$ are classified as elliptical, lenticular, 
and spiral galaxies, respectively.  \citet{kwg93} and \citet{bcf96} 
showed that this method of type classification well reproduces the 
observed type mix.

Luminosity profiles of galaxies are assumed to obey the de Vaucouleurs'
1/4 law for E/S0 galaxies and the exponential law for spirals.  These
profiles are used to compute the isophotal magnitude, which is obtained
by integrating light from the galactic center to a radius at which the
surface brightness reaches a threshold value.  If the threshold is
smaller in units of surface brightness, e.g., mag arcsec$^{-2}$, the
contribution to the isophotal magnitude comes only from the central
region of the galaxy.  Where the threshold is infinity, the isophotal
magnitude is identical to the total magnitude.  Thus, the threshold
depends on the depth of observations.  It is very important for high
redshift galaxies to estimate the isophotal magnitude because of the
selection effects for detection.  The details of how to compute the
isophotal magnitude are shown in \citet{y93}, \citet{ty00} and
\citet{t01}, and in \citet{ntgy01, nytg02} in the framework of SA
models.

\section{Model Parameters}

Here we briefly summarize model parameters.  As already mentioned, we
adopt a standard $\Lambda$CDM model.  The cosmological parameters are
$\Omega_{0}=0.3, \Omega_{\Lambda}=0.7, h=0.7$ and $\sigma_{8}=0.9$, and
the baryon density parameter is $\Omega_{\rm b}=0.048$, as tabulated in
Table \ref{tab:cosmo}, suggested by the first-year {\it WMAP} results
\citep{s03}.  The shape of the power spectrum of density fluctuations
adopted here is that of \citet{s95}, in which the effects of baryons are
taken into account to improve the power spectrum given by \citet{bbks}.

Most of the astrophysical parameters are constrained from local
observations, according to the procedure discussed in \citet{ntgy01,
nytg02} and NY04.  The values adopted are slightly different from those
in our previous papers.  This is mainly caused by different merger trees
which are directly constructed in this paper using $N$-body simulations.
We shall show that correctly constructing merger trees is very important
in \S\S \ref{sec:lf}.  In the Mitaka model (NY04), we used a mass
function which fits with results of the $N$-body simulations we used
here \citep{yny04}, which means that mass functions of dark halos are
the same at output redshift, e.g. at $z=0$.  However, we have found that
luminosity functions in this model are different from those in NY04, in
which an extended Press-Schechter model was used to construct merger
trees (see Figure \ref{fig:lf3}).  This shows that merger trees of dark
halos directly affect properties of galaxies.

Parameters are determined in the following way: Firstly, the SN
feedback-related parameters, $\alpha_{\rm hot}$ and $V_{\rm hot}$ are
constrained by matching luminosity functions (see Figure \ref{fig:lf1}).
Roughly speaking, $\alpha_{\rm hot}$ determines the faint-end slope and
$V_{\rm hot}$ the magnitude of $L^{*}$ galaxies.  Since observational
uncertainties in estimating luminosity functions still remain, we have a
degree of freedom in varying those parameters.  In the Mitaka model, we
have used $\alpha_{\rm hot}=4$, a parameter also used in this paper and
referred to as the SFB model.  In many other SA models, however,
$\alpha_{\rm hot}=2$ is often used \citep[e.g.][]{kwg93, sp99, clbf00}.
Therefore, we provide another parameter set, the WFB model, with
$\alpha_{\rm hot}=2$.  $V_{\rm hot}$ is determined for luminosity
functions to be consistent with observations under the constraint of
$\alpha_{\rm hot}$.  Second, SFR-related parameters, $\tau_{*}^{0}$ and
$\alpha_{*}$, are determined by matching cold gas fractions in spiral
galaxies, $M_{\rm HI}/L_{B}$ (see Figure \ref{fig:cold}).  As mentioned,
we have found that $\alpha_{*}=-\alpha_{\rm hot}$ for the SFB model, but
$\alpha_{*}=-2.3$ for the WFB model, which is slightly different from
$-\alpha_{\rm hot}=-2$.  Finally, other parameters such as the mean spin
parameter are fine-tuned, but they are quite similar to physically
motivated values.  The values of these parameters are tabulated in Table
\ref{tab:astro}.

\begin{figure*}
\plotone{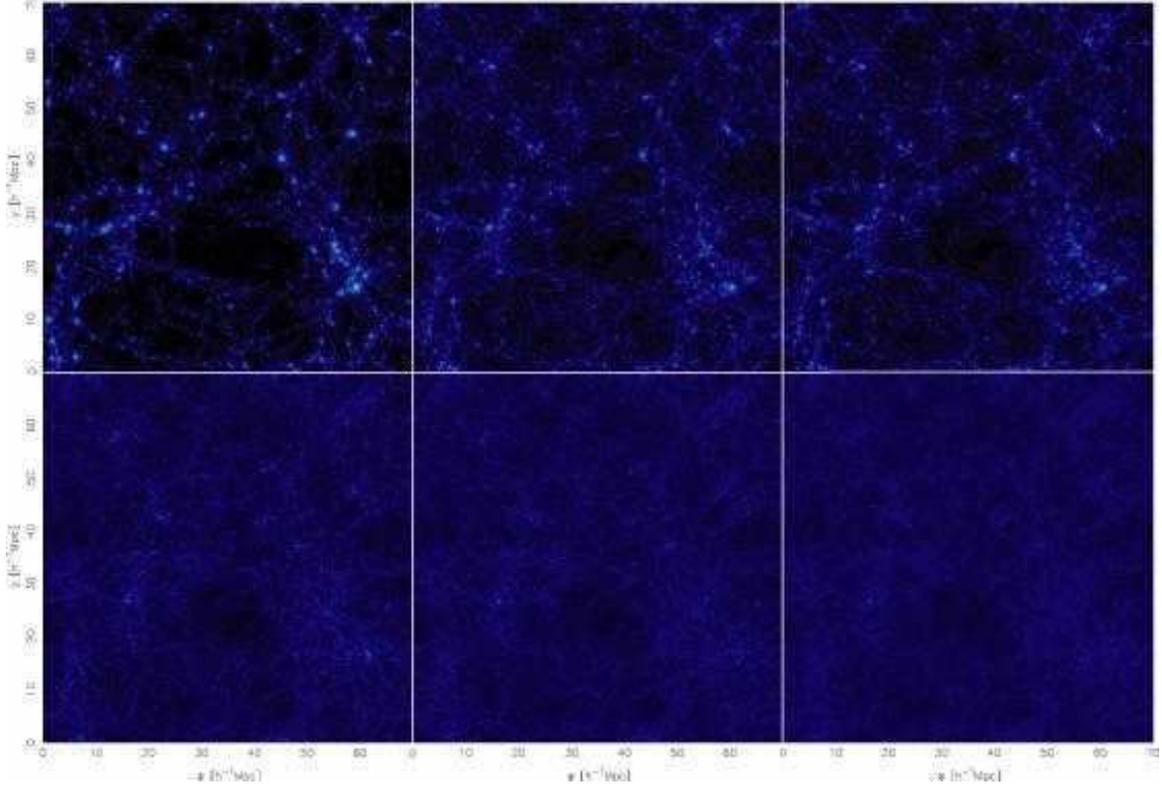}

\caption{Projected distribution of dark matter in the $L=70 h^{-1}$ Mpc
simulation at six redshifts: $z=0$ (top left), 1 (top middle), 2 (top
right), 3 (bottom left), 4 (bottom middle) and 5 (bottom right).
Brighter regions indicate denser regions.  }

\label{fig:dz0}
\end{figure*}

\begin{figure*}
\plotone{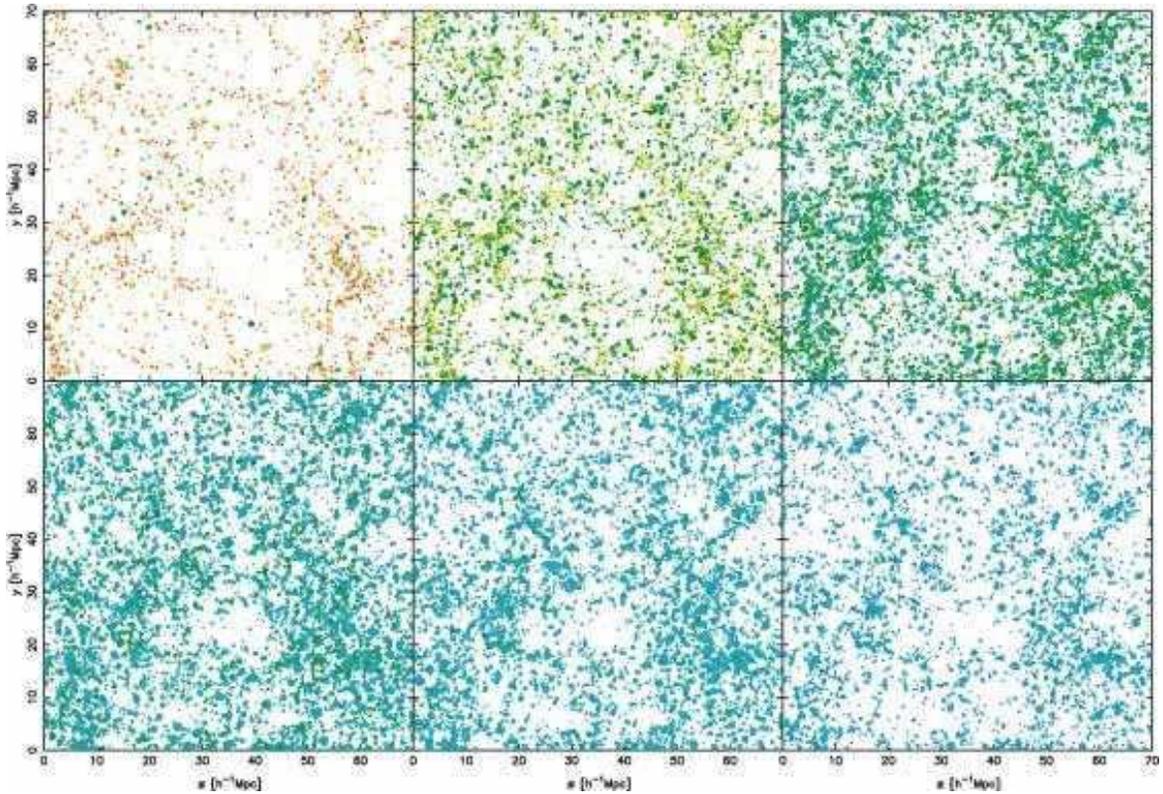}

\caption{Same as Figure \ref{fig:dz0} but for galaxies.  Plotted
galaxies are $M_{B}\leq -19.5$ in their rest frames.  Larger points
represent brighter galaxies.  Colors of points indicate the rest frame
$B-V$ color of galaxies.}

\label{fig:gz0}
\end{figure*}

Figure \ref{fig:dz0} shows the projected density distribution of dark
matter taken from the $L=70 h^{-1}$ Mpc simulation at six different
redshifts from $z=0$ to 5.  It is evident that dark matter is clustering
more as time passes.  Figure \ref{fig:gz0} shows the same distribution
as Figure \ref{fig:dz0} but for galaxies of $M_{B}\leq -19.5$
($M_{B}-5\log h\la -18.7$) in the rest frame.  The position of galaxies
is specified by the marker particles, as mentioned in \S\S \ref{sec:mh}.
Sizes and colors of symbols represent the rest frame magnitudes and
colors of galaxies, respectively.  In this plot, we do not take into
account the selection effects and absorption of light by intergalactic
\ion{H}{1} clouds.  At higher redshifts, galaxies become bluer.  By
contrast, clustering properties of galaxies seem to be almost
independent of redshifts.  This has already been described as ``biased''
galaxy formation, and \citet{kcdw99b} have shown that the correlation
length at which the two-point correlation function of galaxies becomes
unity does not significantly evolve as dark matter does and that the
bias of galaxy distribution to dark matter grows toward higher redshift.
In a subsequent paper, we shall investigate clustering properties of
galaxies in detail \citep{ynegy}.

\section{Galaxies at present}

\subsection{luminosity functions}\label{sec:lf}

\begin{figure}
%\epsscale{0.7}
\plotone{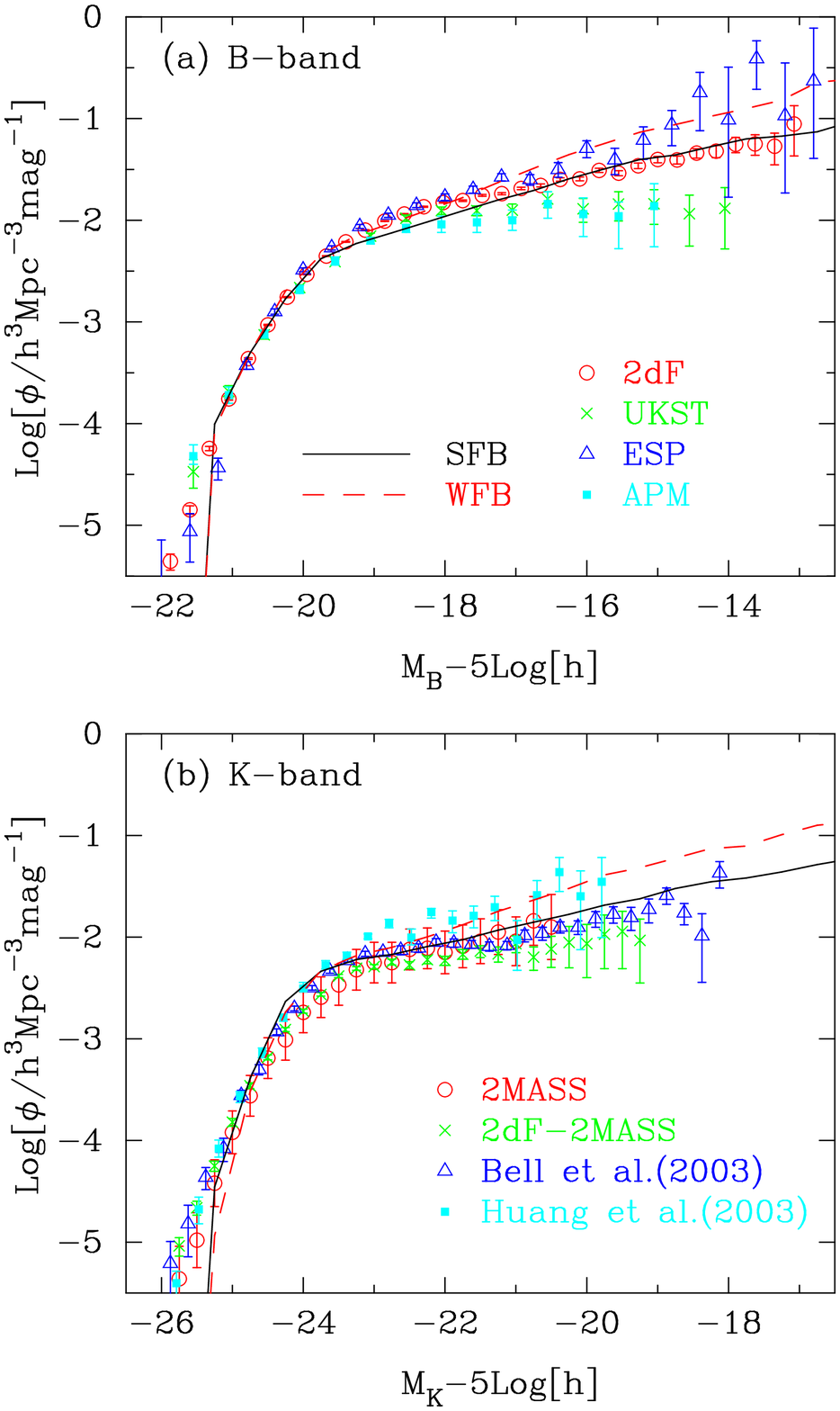}

\caption{Local luminosity functions in the (a) $B$ and (b) $K$ bands.
 The solid and dashed lines represent the SFB and WFB models,
 respectively.  Symbols with error bars in (a) indicate the
 observational data from APM \citep[][{\it filled squares}]{l92}, ESP
 \citep[][{\it open triangles}]{z97}, Durham/UKST \citep[][{\it
 crosses}]{r98}, and 2dF \citep[][{\it open circles}]{f99}.  Symbols in
 (b) indicate the data from 2MASS \citep[][{\it open circles}]{k01}, 2dF
 combined with 2MASS \citep[][{\it crosses}]{c01}, \citet[][{\it open
 triangles}]{bmkw03}, and \citet[][{\it filled squares}]{hgct03}.}

\label{fig:lf1}
\end{figure}

Figure \ref{fig:lf1} shows $B$ and $K$-band luminosity functions of
galaxies.  The solid and dashed lines represent the SFB and WFB models,
respectively.  Since the SF time-scale hardly affects local luminosity
functions, they are determined almost solely by the SN feedback-related
parameters, $\alpha_{\rm hot}$ and $V_{\rm hot}$.  Symbols with error
bars represent observational results from the $B$-band redshift surveys,
such as Automatic Plate Machine \citep[APM;][]{l92}, ESO Slice Project
\citep[ESP;][]{z97}, Durham/United Kingdom Schmidt Telescope
\citep[UKST;][]{r98} and Two-Degree Field \citep[2dF;][]{f99}, and from
the $K$-band redshift surveys given by Two Micron All Sky Survey
\citep[2MASS;][]{k01}, 2dF combined with 2MASS \citep{c01},
\citet{bmkw03} and \citet{hgct03}.  The SFB model agrees well with the
luminosity functions of 2dF and \citet{bmkw03}, and the WFB model agrees
with those of ESP and \citet{hgct03}.

\begin{figure}
%\epsscale{0.7}
\plotone{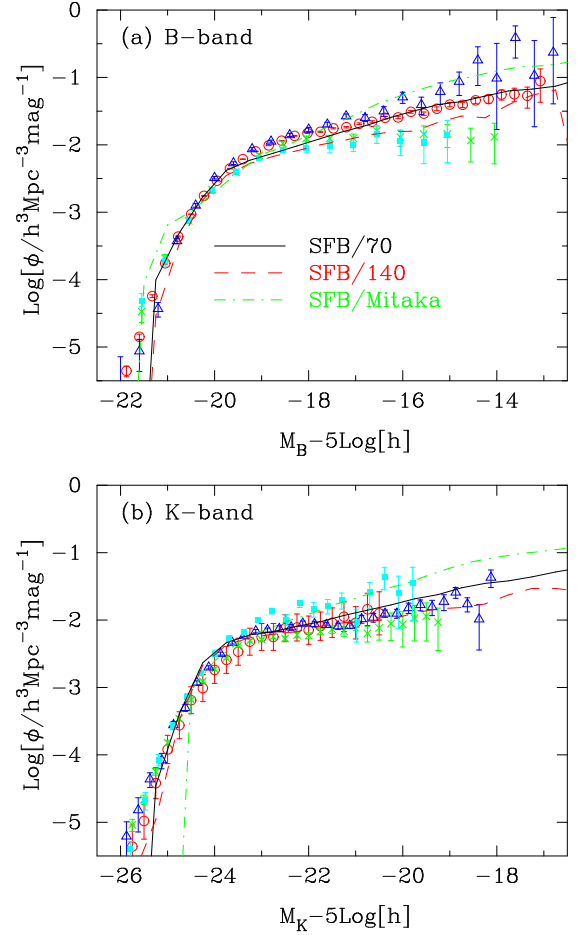}

\caption{Local luminosity functions.  Same as Figure \ref{fig:lf1}, but
for theoretical models plotted.  The solid lines represent the same SFB
model as shown in Figure \ref{fig:lf1}, indicated by SFB/70 in (a).  The
dashed lines represent a model with the same parameters as the SFB
model, but for the simulation box size of 140 $h^{-1}$ Mpc, which has an
eight times larger minimum mass of dark halos.  The dot-dashed lines
represent the Mitaka model given by NY04, but for the same parameters as
the SFB model.  The minimum mass of dark halos in this model is given by
$V_{\rm circ}=40$ km s$^{-1}$.}

\label{fig:lf2}
\end{figure}

It is interesting and important to assess the effects of numerical
resolution on galaxies.  This corresponds to the lower limit of the mass
of dark halos.  The $N$-body simulation we use is of a box size of 70
$h^{-1}$ Mpc.  Since all halos must have at least ten dark matter
particles, the minimum mass $M_{70}$ corresponds to 3.04$\times 10^{9}
M_{\odot}$.  Here we use another result of an $N$-body simulation with a
140 $h^{-1}$ Mpc box and the same number of dark matter particles.  The
minimum mass of halos is $M_{140}=2.43\times 10^{10} M_{\odot}=8M_{70}$.
The resultant luminosity functions for the 140 $h^{-1}$ Mpc box
simulation are shown in Figure \ref{fig:lf2} for the SFB model,
indicated by the dashed lines ({\it SFB/140}), with the same
astrophysical parameters as those of the 70 $h^{-1}$ Mpc box simulation
({\it SFB/70; solid lines}).  It is evident that this predicts fewer
galaxies and a very shallow slope at the faint end.  One of the reasons
is clearly that the minimum mass of dark halos becomes larger.  This is
an effect similar to that resulting from the adoption of stronger SN
feedback for low-mass halos so that hot gas cannot cool within dark
halos below $M_{140}$.  This means that if one would like to connect an
obtained parameter set with physical processes such as the SN feedback,
the numerical resolution would be crucial, while we should be able to
construct a predictable model regarding those parameters as only
phenomenological.  Also plotted by the dot-dashed lines is the Mitaka
model, but with the same parameters as those of the SFB model, which are
slightly different from the original ones.  In this model, the minimum
mass of halos corresponds to $V_{\rm circ}=40$ km s$^{-1}$.  Part of the
reason the faint-end slope is steeper than that of the SFB model should
be that the minimum mass becomes smaller at high redshift when stars
actively form.  This is, however, only part of the reason.  This is
shown below.  From this comparison, we would like to stress that we
should denote attention to the minimum mass of dark halos, or the
numerical resolution, which corresponds physically to the effective
Jeans mass.

\begin{figure}
%\epsscale{0.7}
\plotone{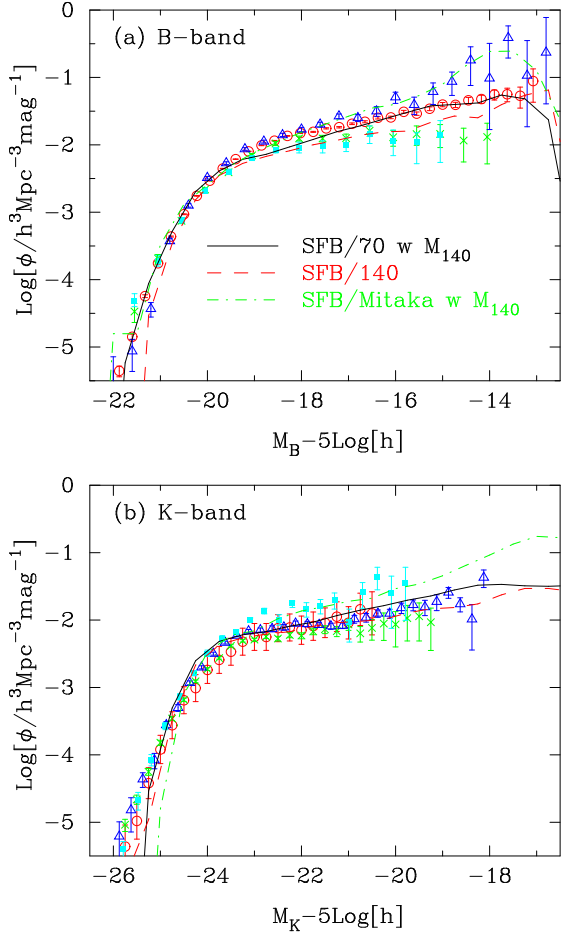}

\caption{Local luminosity functions.  Same as Figure \ref{fig:lf1}, but
for theoretical models.  The solid lines represent the SFB model with
the same minimum mass $M_{140}$ as the SFB/140 model.  Hot gas within
halos between $M_{70}$ and $M_{140}$ do not cool at all.  The dashed
lines represent the same SFB/140 model as shown in Figure \ref{fig:lf2}.
The dot-dashed lines represent the Mitaka model with the minimum mass
$M_{140}$.}

\label{fig:lf3}
\end{figure}

Further investigation of this problem provides interesting features.
Figure \ref{fig:lf3} shows luminosity functions for the SFB and Mitaka
models with the minimum mass $M_{140}$ of halos indicated by the solid
and dot-dashed lines, respectively, as well as the SFB/140 model by the
dashed lines.  At first, it is shown that the SFB/70 model with
$M_{140}$ does not recover the SFB/140 model, except for the cut-off at
the faint end, $M_{B}-5\log h\sim -14$.  The change in the slope caused
by the change in minimum mass is quite small.  Therefore, the minimum
mass seems to mainly affect only very faint galaxies.  Second, the
faint-end slope of the Mitaka model with $M_{140}$ also hardly changes
except for the cut-off.  The bump at $M_{B}-5\log(h)\simeq -14$ is a
characteristic luminosity of the faintest galaxies determined by the
higher minimum mass.  In other words, galaxies formed in dark halos of
the minimum mass without any progenitors should have such luminosities
under the given strength of the SN feedback.  Thus it is difficult for
those to have fainter luminosities.  This figure therefore shows again
that the numerical resolution significantly affects galaxy formation,
and that there is a limitation in the extended Press-Schechter formalism
which gives us merging histories of dark halos.

We would like to stress here that this is {\it not} a convergence test.
If the merger trees included less massive dark halos, the results would
change further.  In this case, the faint end of the luminosity function
would become steeper because such small halos produce very faint
galaxies.  In our current model, although no realistic model of the
effective Jeans mass is included, the mass resolution corresponding to
$M_{70}$ should be required for realistic modeling, because adopting
$M_{140}$ or larger minimum mass corresponds to overestimation of the
effective Jeans mass.  It should be very important to model the
effective Jeans mass by utilizing higher-resolution simulations in
future.

\begin{figure}
%\epsscale{0.7}
\plotone{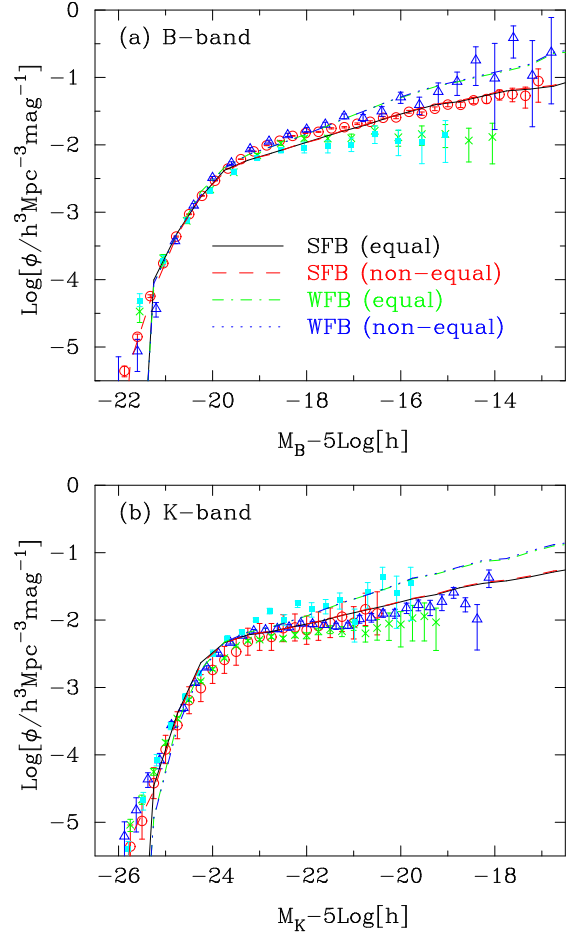}

\caption{Local luminosity functions.  Same as Figure \ref{fig:lf1}, but
for two added variant models.  The solid and dot-dashed lines represent
the SFB and WFB models, respectively, the same as those shown in Figure
\ref{fig:lf1}.  The dashed and dotted lines also represent the SFB and
WFB models, but including the effect of non-equal mass mergers between
satellite galaxies described by eq.(\ref{eqn:mamon2}).  Note that
luminosity functions of models with and without the effect of non-equal
mass mergers are almost identical.}

\label{fig:lf4}
\end{figure}

Finally, we discuss mergers of satellite galaxies.  As shown in
eq.(\ref{eqn:mamon2}), our code enables us to take into account the
effect of non-equal mass mergers between satellite galaxies based on
\citet{m00}, while this is normally switched off.  Figure \ref{fig:lf4}
shows this effect on luminosity functions.  The dashed and dotted lines
indicate the SFB and WFB models with the effect of non-equal mass
mergers, while the solid and dot-dashed lines indicate the SFB and WFB
models without this effect, which are the same as those shown in Figure
\ref{fig:lf1}.  It is easily seen that the luminosity functions with and
without the effect of non-equal mergers are almost identical except for
the bright end.  Therefore, we conclude that, fortunately, we do not
need to consider the effect of non-equal mass mergers.  Thus we consider
only equal mass mergers using the rate of $k_{\rm MH}$ given by
\citet{mh97} below.

\subsection{cold gas}

The amount of cold gas, which corresponds to the interstellar medium, is
mainly determined by SN feedback-related and SFR-related parameters.
The former parameters determine the gas fraction expelled from galaxies,
and the latter, the gas fraction that is converted into stars.  Because
the SN feedback-related parameters are already constrained by matching
luminosity functions as shown in the previous subsection, we can use
observed gas fractions in spiral galaxies as constraints on the
SFR-related parameters, $\alpha_{*}$ and $\tau_{*}^{0}$.

\begin{figure}
\plotone{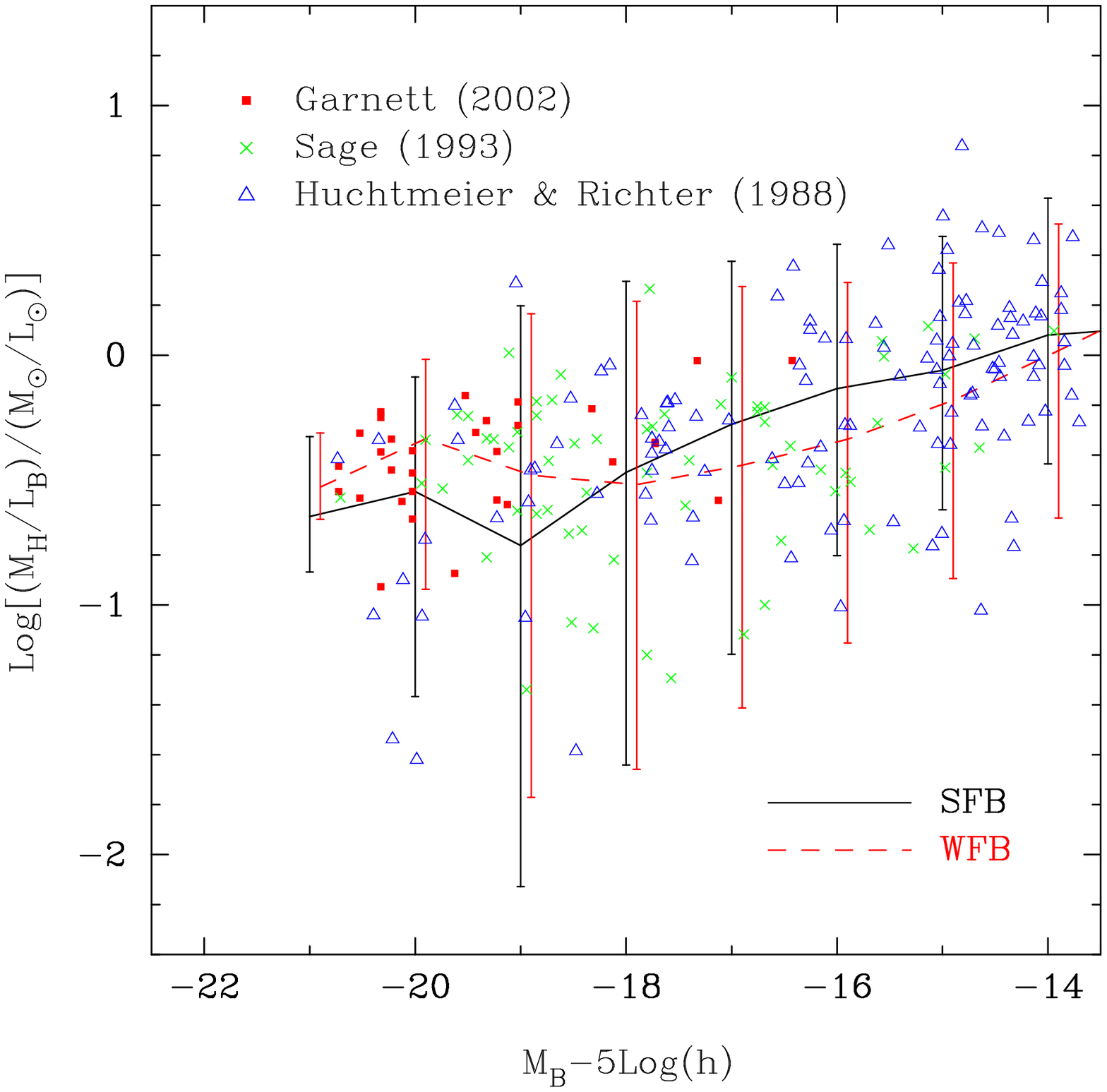}

\caption{Hydrogen mass relative to $B$-band luminosity of spiral
galaxies.  The solid and dashed lines represent the medians for the SFB
and WFB models, respectively, of the distributions of $M_{\rm H}/L_{B}$.
The error bars indicate 25\% and 75\% percentile points.  Symbols
indicate the observational data from \citet[][{\it open
triangles}]{hr88}, \citet[][{\it crosses}]{sage93}, and \citet[][{\it
solid squares}]{garnet02}.  Note that the data from \citet{hr88} include
only \ion{H}{1}.}

\label{fig:cold}
\end{figure}

Figure \ref{fig:cold} shows the ratio of cold gas mass relative to
$B$-band luminosity of spiral galaxies as a function of their
luminosity.  We assume here that 75\% of the cold gas in the models is
comprised of hydrogen, i.e., $M_{\rm H}=0.75M_{\rm cold}$.  The solid
and dashed lines associated with error bars represent the medians and
25\% and 75\% percentiles of the distributions for the SFB and WFB
models, respectively.  We have added 0.1 mag offsets to the data of the
WFB model for error bars to be distinguishable.  Symbols indicate the
observational data of individual galaxies taken from \citet[][{\it open
triangles}]{hr88}, \citet[][{\it crosses}]{sage93}, and \citet[][{\it
solid squares}]{garnet02}.  Note that the data from \citet{hr88} include
only \ion{H}{1}, and that the fraction of hydrogen molecule, H$_{2}$,
usually increases with the luminosity.  Although the medians for both
models are in good agreement with the observations, the error bars for
the theoretical models seem to be larger than the distribution of the
observational data.  This might suggest the existence of
self-regularization mechanisms of star formation to develop small
scatters of cold gas fraction.  Nevertheless, we can say that the models
agree well with the observational data.

Adopted values of the parameters for the SF model are tabulated in Table
\ref{tab:astro}.  The SFB model has $\alpha_{*}=-4$, which means
$\tau_{*}\propto \beta$.  This is similar to the prescription of the SF
time-scale in NY04, $\tau_{*}\propto(1+\beta)$.  On the other hand, in
the WFB model, we have to set $\alpha_{*}=-2.3\neq -\alpha_{\rm hot}$.
Unfortunately, the simple prescription of NY04 does not work for weak SN
feedback models.

\begin{figure}
\plotone{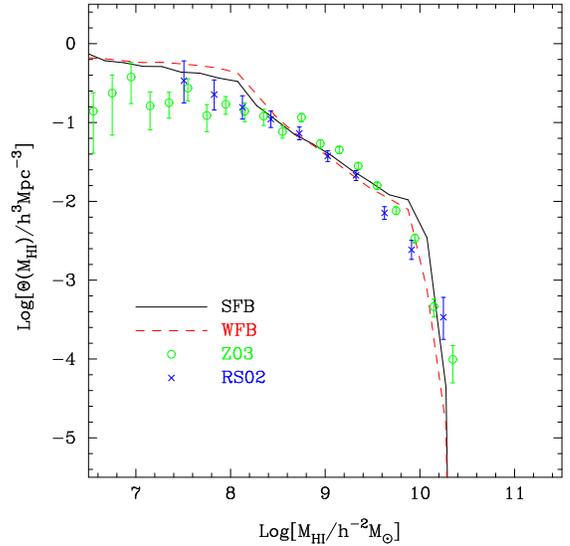}

\caption{\ion{H}{1} mass functions.  The solid and dashed lines
 represent the SFB and WFB models, respectively.  The crosses and open
 circles indicate the observational data from \citet{rs02} and
 \citet{z03}, respectively.  Again we set $M_{HI}=0.75M_{\rm cold}$ for
 the theoretical models.}

\label{fig:HIMF}
\end{figure}

Recent radio surveys have provided good data of \ion{H}{1} mass
functions (HIMFs).  Figure \ref{fig:HIMF} shows the theoretical
prediction of HIMFs for the SFB and WFB models, respectively.  Both
models provide almost identical HIMFs, because we have chosen the
SFR-related parameters to be consistent with the observed cold gas
fractions.  Symbols indicate the observational data from \citet[][{\it
crosses}]{rs02} and \citet[][{\it open circles}]{z03}.  Both models
reproduce the observed HIMFs quite well.  Note that we again assume
$M_{\rm HI}=0.75M_{\rm cold}$.

\subsection{galaxy sizes}

\begin{figure}
\plotone{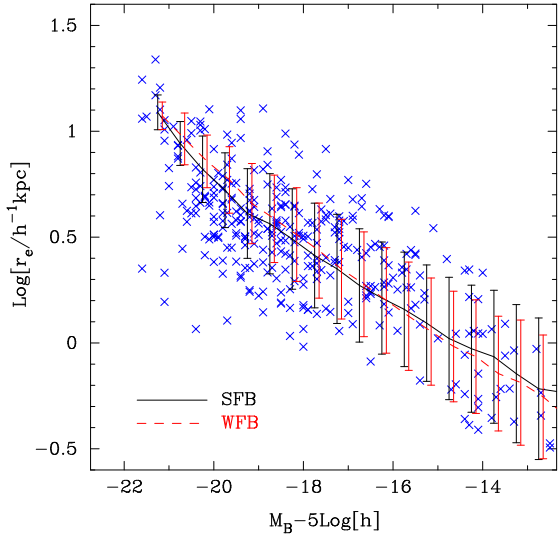}

\caption{Disk size of spiral galaxies.  The solid and dashed lines
represent the averages of disk radii of spiral galaxies in the SFB and
WFB models, respectively.  Error bars on these lines are 1 $\sigma$
scatter in predicted sizes.  To avoid any confusion, 0.1 mag offsets are
imposed on the results of the WFB model.  Crosses indicate the
observational data taken from \citet{isib96}.}

\label{fig:r}
\end{figure}

Figure \ref{fig:r} shows the effective disk radii of local spiral
galaxies as a function of their luminosities.  The solid and dashed
lines represent the SFB and WFB models, respectively.  Symbols are the
observational data taken from \citet{isib96}.  To avoid confusion, we
have added 0.1 mag offsets to the WFB model.  Here we would like to
briefly repeat the procedure to determine the disk size.  Dark halos
form with angular momenta randomly assigned with the log-normal
distribution of the dimensionless spin parameter, $\lambda_{\rm H}$
(eq. \ref{eqn:lognormal}).  Hot gas contained in a dark halo is assumed
to have the same specific angular momentum as that of the halo.  The hot
gas cools and contracts until the cooled gas is rotationally supported.
Thereby the size of disks is determined.  Good agreement of the
theoretical models with the observational data shown in this figure
means that the standard assumption we adopt for determining disk sizes
works well when the distribution of the dimensionless spin parameter is
in a reasonable range, estimated from the structure formation theory in
the CDM universe.

Correctly estimating disk sizes is very important to compare theoretical
models with observations because it affects many observational
properties of galaxies such as the surface brightness and the dust
column density.  \citet{ntgy01, nytg02} have discussed such effects
caused by the uncertainty in estimating disk sizes.

\begin{figure}
%\epsscale{0.7}
\plotone{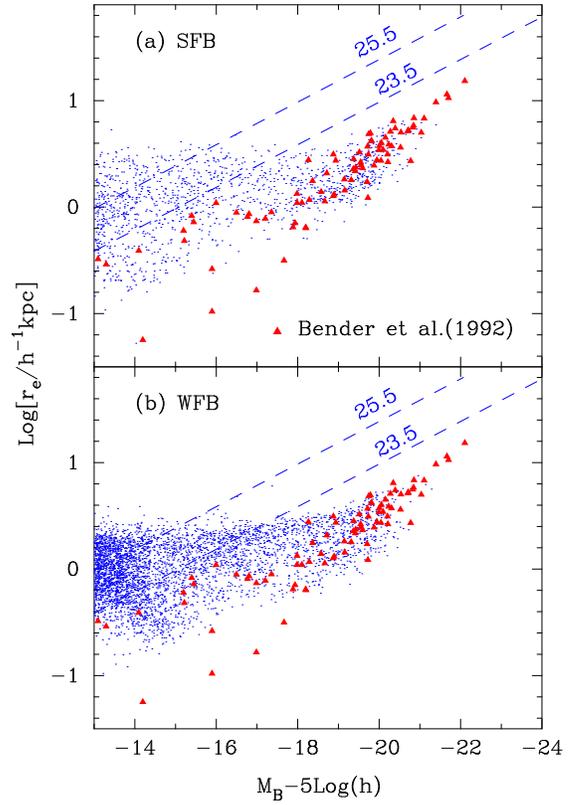}

\caption{Sizes of elliptical galaxies.  The dots represent effective
radii of individual elliptical galaxies in (a) the SFB model and (b) the
WFB model.  Only a quarter of all elliptical galaxies is plotted.  The
filled triangles indicate the observational data taken from
\citet{bbf92}.  The dashed lines denote constant surface brightnesses,
23.5 and 25.5 mag arcsec$^{-2}$, as indicated in panels.}

\label{fig:rad}
\end{figure}

Figure \ref{fig:rad} shows the effective radii of elliptical galaxies
against their $B$-band magnitudes.  The dots represent individual
elliptical galaxies in the (a) SFB and (b) WFB models.  The effective
radius $r_{e}$ is estimated from $r_{e}=0.744 r_{b}$ \citep{ny03}, where
$r_{b}$ is the three-dimensional half-mass radius [eqs.(\ref{eqn:sizeE})
and (\ref{eqn:sizeE2})].  To make figures clearer, only a quarter of all
ellipticals are plotted.  The solid triangles indicate the observational
data taken from \citet{bbf92}.  The theoretical models agree well with
the observational data of at least brighter than $M_{B}-5\log(h)\simeq
-18$.  At magnitudes fainter than that, the fraction of compact
ellipticals in the models seems to be smaller than that expected from
the observations.  This would be partly caused by the limited
resolution.  Another possibility is that the SF time-scale at high
redshift might be slightly too long.  As clearly shown by NY04, the size
of dwarf ellipticals strongly depends on the SF time-scale at high
redshift.  For example, let us consider the SF time-scale simply
proportional to the dynamical time-scale.  In this case, the SF
time-scale is shortened for higher redshift, and at $z=2$, it becomes
nearly a quarter of that at $z=0$.  Noting that the SF time-scale also
determines the rate of gas consumption, therefore, galaxies are
statistically gas-poorer than those in the case of the constant SF
time-scale over all redshifts at high redshift, at which galaxies
frequently merge.  Evidently, because the change in size and velocity
dispersion due to the dynamical response becomes smaller, galaxies with
small sizes and large velocity dispersions increase.  As shown by NY04,
some compact ellipticals represented in Figure \ref{fig:rad} are closely
related to ellipticals formed via gas-poor mergers.  Although we are
able to make such compact ellipticals by slightly shortening the SF
time-scale at high redshift with an extra parameter relating to the
redshift dependence, we would like to leave it constant to avoid further
complexity of the model.  A detailed discussion about the relationship
between the SF time-scale and the size and velocity dispersion of
ellipticals can be found in NY04.

The dashed lines represent constant surface brightnesses of 23.5 and
25.5 mag arcsec$^{-2}$ as indicated in the figure.  Because the
detection of galaxies is limited by the surface brightness, galaxies
with surface brightness lower than the detection limit, which depends on
observations, cannot be detected.  Although our models apparently
predict too many dwarf ellipticals with radii larger than about 1
$h^{-1}$ kpc, they will not be observed unless ultra deep observations
are performed.  Note that the samples of galaxies from \citet{bbf92} may
be expected to have sufficiently high surface brightness to investigate
their structural and kinematical properties, because the purpose of the
paper is to clarify those properties.  Thus it should not be a problem
that many dwarf ellipticals with much larger radii emerge.  In fact, the
selection effects brought about by the surface brightness play a
significant role for high redshift galaxies.  This will be shown in
\S\ref{sec:highz}.

\subsection{velocity-magnitude relations}

Here we investigate velocity-magnitude relations, that is, the
Tully-Fisher (TF) relation for spiral galaxies \citep{tf77} and the
Faber-Jackson (FB) relation for elliptical galaxies \citep{fj76}.

\begin{figure}
\plotone{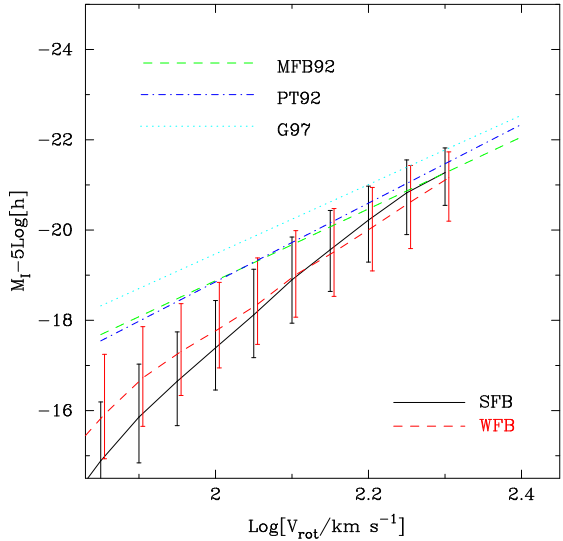}

\caption{$I$-band Tully-Fisher relation of spiral galaxies.  The solid
and dashed lines associated with error bars indicate the medians and
25\% and 75\% points of the distributions of $I$-band magnitudes for the
SFB and WFB models, respectively.  To avoid confusion, 0.1 mag offsets
are imposed on the results of the WFB model.  The observed TF relations
are shown as a mean relation given by \citet[][{\it thin dashed
line}]{mfb92}, \citet[][{\it thin dot-dashed line}]{pt92}, and
\citet[][{\it thin dotted line}]{g97}.  }

\label{fig:tf}
\end{figure}

Figure \ref{fig:tf} shows the $I$-band TF relation.  The solid and
dashed lines associated with error bars represent the medians and 25\%
and 75\% points of the distributions of $I$-band magnitudes for the SFB
and WFB models, respectively.  While gas-rich spirals tend to be more
luminous as indicated by \citet{clbf00} and NY04, we show only the
results for all spiral galaxies.  We find that, although not shown, the
TF relation produced only by gas-rich spirals becomes too bright,
particularly at $V_{\rm rot}\sim 130$ km s$^{-1}$ and too steep at
$V_{\rm rot}\la 100$ km s$^{-1}$ to reconcile with observations in both
models.  The straight lines indicate the results best fit to the
observed TF relations of \citet[][{\it dashed}]{mfb92}, \citet[][{\it
dot-dashed}]{pt92}, and \citet[][{\it dotted}]{g97}.  We assume that the
linewidth $W$ is simply twice the disk rotation velocity, $V_{\rm rot}$,
as usual.  For model galaxies, we set $V_{\rm rot}=V_{d}$.

At faster than $V_{\rm rot}\simeq 160$ km s$^{-1}$ [$\log(V_{\rm
rot}/{\rm km~ s}^{-1})\simeq 2.2$], both the theoretical models agree
well with the observational TF relations.  Besides, the WFB model is
marginally consistent with some of the observations down to $V_{\rm
rot}\simeq 100$ km s$^{-1}$, while the TF relation of the SFB model
deviates and has a steeper slope compared with the observations.  As
shown in NY04, the slope is mainly determined by $\alpha_{\rm hot}$
because it provides the dependence of the strength of the SN feedback on
the velocity and therefore affects luminosities, particularly of dwarf
galaxies.  The deviation at small velocities might disappear if the
dynamical response on the rotation velocity to gas removal due to the SN
feedback is taken into account, which is currently applied only to
spheroidal formation by starbursts.  Because dwarf galaxies
significantly suffer from the SN feedback, disks may expand and rotate
slowly.

\begin{figure}
%\epsscale{0.7}
\plotone{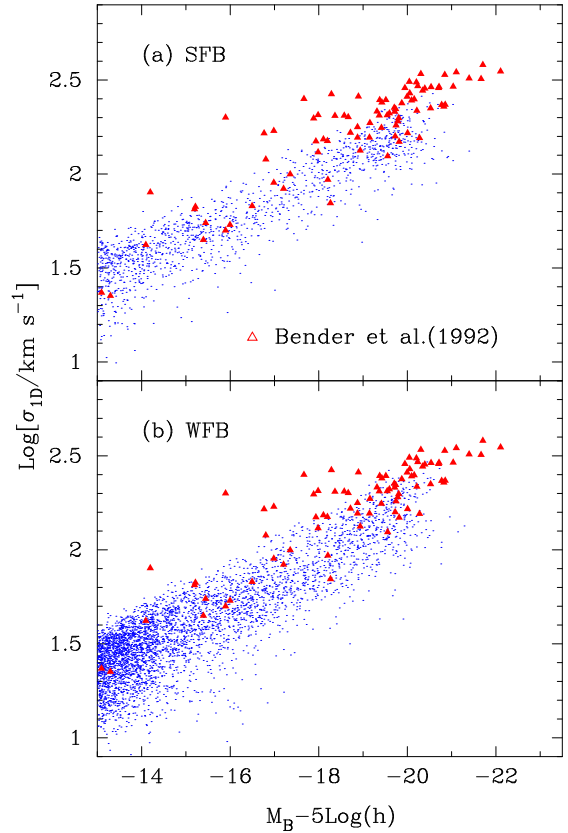}

\caption{$B$-band Faber-Jackson relation of elliptical galaxies.  The
dots represent one-dimensional velocity dispersions of individual
elliptical galaxies in (a) the SFB model and (b) the WFB model.  Only a
quarter of all elliptical galaxies is plotted.  The filled triangles
indicate the observational data taken from \citet{bbf92}.  The velocity
dispersion of the models is simply estimated as
$\sigma_{1D}=V_{b}/\sqrt{3}$.}

\label{fig:fj}
\end{figure}

Figure \ref{fig:fj} shows the $B$-band FJ relation of elliptical
galaxies.  The dots represent individual ellipticals in the (a) SFB and
(b) WFB models.  The model predictions broadly agree with the
observational data from \citet{bbf92}.  This agreement is obtained
thanks to $f_{\rm diss}=1.3$.  In the case of $f_{\rm diss}=1$ as in
NY04, the models predict smaller velocity dispersions than the
observations.  The reason $f_{\rm diss}=1.3$ is required in this model
would be the differences of the SN feedback and the SF time-scale.
Dwarf galaxies with $\sigma_{\rm 1D}\la 20$ km s$^{-1}$ can be obtained
only by associating with the dynamical response (see Figure
\ref{fig:vm}) as shown by NY04, in which the effect of the dynamical
response was introduced into SA models for the first time.  Detailed
discussion on the velocity dispersion of dwarf galaxies can be found in
the previous subsection and NY04.

\subsection{color-magnitude relations}\label{sec:cmr}

\begin{figure}
%\epsscale{0.7}
\plotone{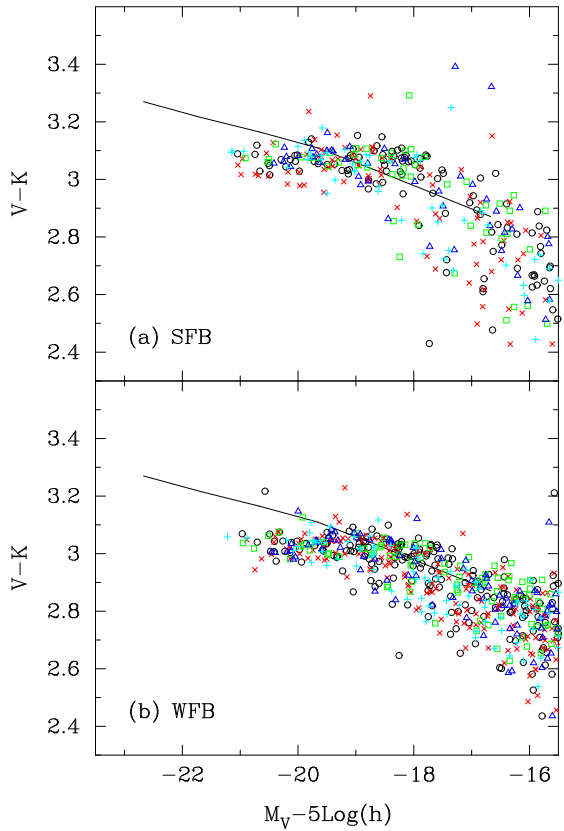}

\caption{$V-K$ color-magnitude relations of elliptical galaxies in five
of the most massive clusters in the (a) SFB and (b) WFB models.
Different symbols represent ellipticals in different clusters.  The
solid line is the observed CM relation of the Coma cluster with a
correction to the aperture effect by \citet{kaba98}}

\label{fig:cmr}
\end{figure}

Figure \ref{fig:cmr} shows the $V-K$ color-magnitude (CM) diagram for
cluster elliptical galaxies.  Cluster halos that we pick out have masses
of 6.18, 4.07, 2.99, 2.71, and 2.68 $\times 10^{14}M_{\odot}$, which are
five of the most massive halos in our simulation.  Different symbols
represent elliptical galaxies in different clusters.  The solid line
indicates a CM relation provided by \citet{kaba98}, which is corrected
for the aperture effect on the observational CM relation for the Coma
cluster obtained by \citet{ble92}.  While the models agree with the
observation for dwarf ellipticals, $M_{V}-5\log(h)\ga -18$, the
predicted CM relations become almost flat for giant ellipticals.  This
trend is similar to previous work on SA models whose model parameters
are chosen so that many other observations such as luminosity functions
are reproduced \citep{clbf00, ny04}.  The mechanism itself to make the
slope of the CM relation is well understood in terms of the SN feedback.
The slope for dwarf ellipticals is mainly determined by $\alpha_{\rm
hot}$ and the break point at which the CM relation becomes flat mainly
by $V_{\rm hot}$ \citep{kc98, ng01}.

To improve this situation, the metal enrichment should be much more
carefully considered because the CM relation mainly follows the
metallicity-luminosity relation \citep{ka97, kc98, ng01}.  In this
paper, the chemical yield is set at twice the solar value
\citep[e.g.][]{kc98, sp99}.  This is required to obtain colors of
cluster ellipticals similar to observations.  However, this value is too
large compared with yields derived from standard IMFs for disk star
formation such as Salpeter's, which provide almost solar values.
Moreover, a recent SA analysis by \citet{no04} has shown that a
Salpeter-like IMF successfully reproduces observed stellar distributions
on an [O/Fe]-[Fe/H] plane.  A possible solution is to adopt different
IMFs for different modes of star formation.  \citet{baugh05} and
\citet{nlbfc05} have suggested Kennicutt's IMF for disk star formation
and a top-heavy IMF for starbursts.  Although this should be worth
considering, it is beyond the scope of this paper at this stage of our
modeling.

\section{High redshift galaxies}\label{sec:highz}

\subsection{Faint Galaxy Number Counts}

\begin{figure*}
\plotone{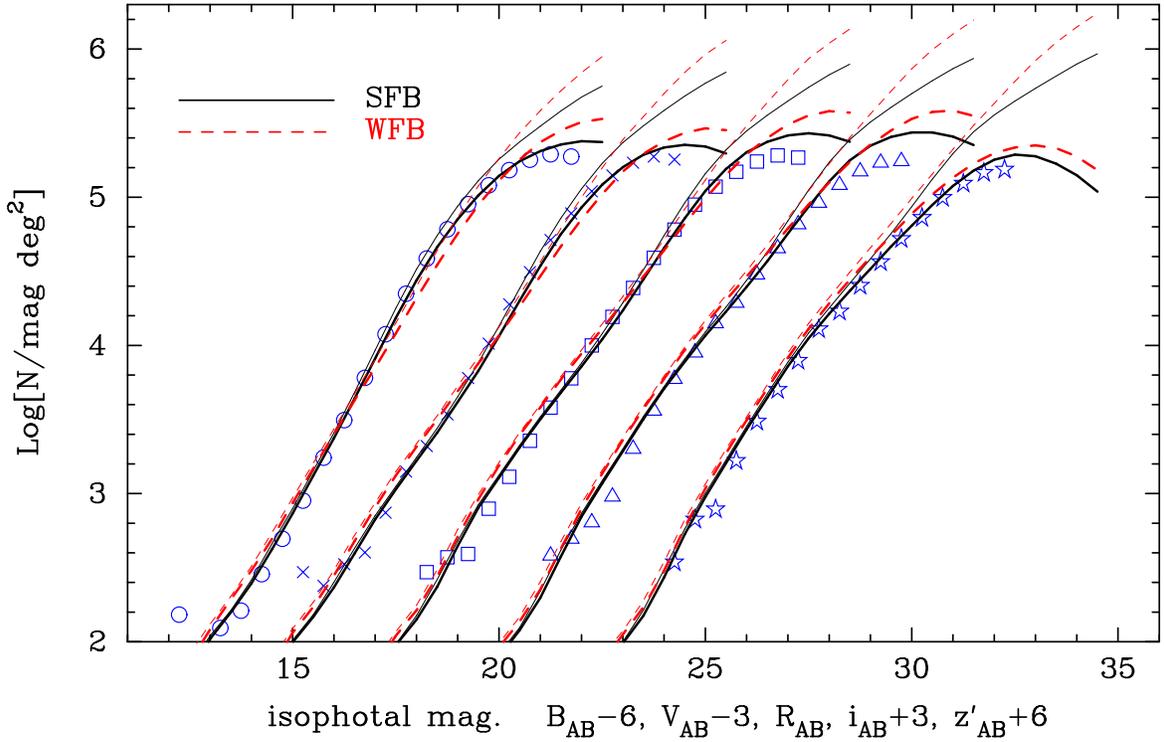}

\caption{Galaxy number counts in the $BVRi'z'$ bands from the left to
the right.  To avoid any confusion, offsets of -6, -3, +3, and +6
magnitudes are added for $B, V, i', $ and $z'$ bands, respectively.  The
solid and dashed lines represent the SFB and WFB models, respectively.
The thick and thin lines denote models with and without the selection
effects, respectively, arising from the same detection threshold of
surface brightness of galaxies as employed in the SDS.  Symbols indicate
the observed SDS counts.}

\label{fig:sdf1}
\end{figure*}

Counting galaxies in a unit area is a task fundamental to observations
in astrophysics \citep[e.g.,][]{yt88}.  It also provides important
constraints on galaxy formation theories as shown in \citet{ntgy01,
nytg02}.  Figure \ref{fig:sdf1} shows faint galaxy number counts in the
optical $BVRi'z'$-bands.  Symbols indicate observed counts taken from
\citet{k04}, obtained by the {\it Subaru Deep Survey} (SDS) using the
Suprime-Cam with a substantially large field-of-view, $34'\times 27'$,
so the data are very reliable in a statistical sense, compared with
observations by other telescopes such as the {\it Hubble Space
Telescope} (HST).  In this figure, we add offsets to magnitudes
depending on pass-bands to avoid confusion.  Note that these data are
observationally {\it raw}, which means that no correction to selection
effects is considered and that the isophotal AB magnitude is used.  The
solid and dashed lines represent the SFB and WFB models, respectively.
The thick and thin lines denote models with and without selection
effects arising from cosmological dimming of surface brightness
\citep{y93}.  To estimate the selection effects, we need detection
limits of surface brightness dependent on observations and on seeing of
observations which determines the FWHM of images.  The details of the
selection effects are shown in \citet{ty00}, and in \citet{ntgy01}
within the framework of SA models.  In these optical pass-bands, as
shown in Figure \ref{fig:sdf1}, both models agree incredibly well with
the observed galaxy counts when the selection effects are included.  We
have confirmed that these models also reproduce galaxy counts well in
the Hubble Deep Field (HDF), as shown for the Mitaka model in NY04.
Note that models of dust optical depth for the SFB and WFB models are
different at high redshift as shown in eq.(\ref{eqn:dust}).  We adopt
$\rho=1$ for the SFB model and $\rho=0$ for the WFB model, which means
that the dust extinction at high redshift in the WFB model is stronger
than in the SFB model.  If we adopted $\rho=0$ for the SFB model, it
would underestimate the number of faint galaxies because of the strong
dust extinction.

\begin{figure}
\plotone{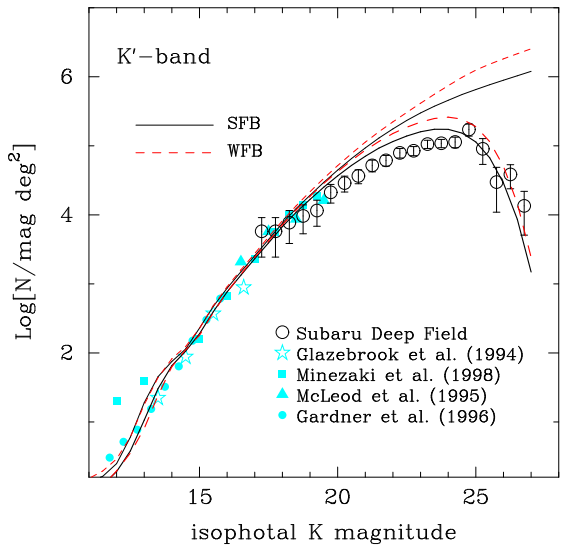}

\caption{Galaxy number counts in the $K'$ band.  Same as Figure
\ref{fig:sdf1}, but for the near-infrared counts in the SDF.  Open
circles associated with error bars indicate the observed SDF counts from
\citet{m01}.  Other symbols are ground-based observed counts as
indicated in the figure.}

\label{fig:sdf2}
\end{figure}

Figure \ref{fig:sdf2} shows galaxy counts in the near-infrared
$K'$-band.  Open circles associated with error bars indicate observed
counts in the {\it Subaru Deep Field} (SDF) taken from \citet{m01}.  We
should warn readers that the field-of-view for the near-infrared SDF is
only $2'\times 2'$, which is less than 1\% of the area for the optical
SDF.  Other symbols are ground-based observed counts from \citet[][{\it
open stars}]{g94}, \citet[][{\it filled squares}]{mkyp98}, \citet[][{\it
filled triangles}]{mbrtf95}, and \citet[][{\it filled circles}]{gscf96}.
The plotted data are also raw counts.  The lines denote the same
meanings as those in Figure \ref{fig:sdf1}.  For $K'$-band counts,
moreover, the completeness arising from noise and statistical
fluctuations is also taken into account.  The details are shown in
\citet{t01} and \citet{nytg02}.  Different from the optical counts shown
in Figure \ref{fig:sdf1}, the agreement with the observations is worse
for both models, which overestimate the number of galaxies at $K'\sim
23$.  The SFB model provides somewhat better agreement than the WFB
model.  This is because, in the WFB model, more galaxies are formed at
high redshift than in the SFB model, due to the weakness of the SN
feedback.  Note that the $K'$ band magnitude mainly reflects the stellar
mass of galaxies, while optical pass-bands magnitudes reflect the SFR.
Because the $K'$ band magnitude is not significantly affected by dust
extinction compared with optical bands, one might consider that a
possible way for the WFB model to recover the observed near-infrared
counts should be stronger selection effects.  However, such a
manipulation also affects the optical galaxy counts, which are already
reproduced.  Thus, in the range of current modeling, we prefer the SFB
model to the WFB, from the $K'$ band galaxy counts.

Finally, we should recall that the above results are due to the choice
of the redshift dependences of the size and optical depth, $\rho$ and
$\gamma$.  In order to obtain the reasonable agreement shown here, it is
necessary to introduce these parameters into simple models of estimating
the disk size and the dust optical depth.  Although it is quite
difficult at this stage to derive the values physically, they will
provide important keys to understanding of the physics of galaxy
formation at high redshift.

\subsection{Redshift distributions}\label{sec:zdist}

\begin{figure}
%\epsscale{0.7}
\plotone{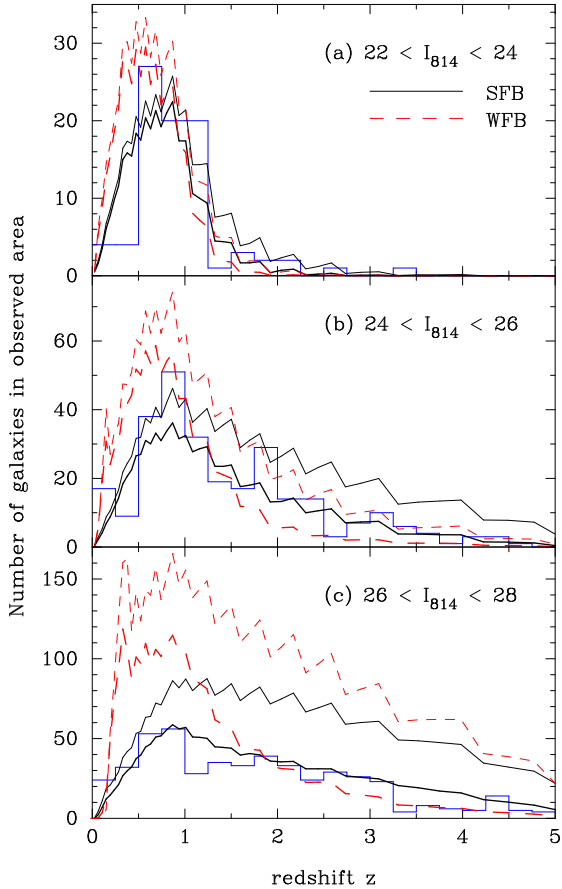}

\caption{Redshift distribution of the HDF galaxies for (a) $22<
I_{814}\leq 24$, (b) $24< I_{814}\leq 26$, and (c) $26< I_{814}\leq 28$.
The solid and dashed lines represent the SFB and WFB models,
respectively.  The thick and thin lines denote models with and without
the selection effects, respectively.  The histogram in each panel is the
observed photometric redshift distribution by \citet{f00}.}

\label{fig:z}
\end{figure}

Figure \ref{fig:z} shows $I_{814}$ band photometric redshift
distribution for the HDF galaxies.  From the top to the bottom, galaxies
for the bright-magnitude bin to the faint-magnitude bin are shown.  The
solid and dashed lines represent the SFB and WFB models, respectively.
The thick and thin lines denote models with and without the selection
effects, which are estimated from the same detection criteria as those
for the HDF.  The histograms are the observed photometric redshift
distribution taken from \citet{f00}.  In the brightest bin (panel {\it
a}), both models agree well with the observation.  The SFB model also
agrees with the observation over all bins.  However, as magnitudes of
galaxies go fainter, the deviation from the observation for the WFB
model increases.  Although the peak of the distribution of galaxies
above the surface brightness threshold for the HDF is similar to the
observation, too many galaxies are detected in the WFB model to be
consistent with the observation.  The weak SN feedback is obviously in
part responsible for this deviation, because it cannot suppress the
formation of dwarf galaxies at low redshift.  Another important reason
is the selection effects.  As shown by the thin lines, the number of
galaxies including undetectable ones is comparable (panel {\it b}) or
sufficiently large (panel {\it c}) compared with the observation.
However, since the selection effects are quite strong, only a small
fraction of galaxies at high redshift can satisfy the detection criteria
for the HDF.  Therefore, the number of detectable galaxies for the WFB
model is smaller than that for the SFB model at high redshift.  Of
course, it is possible to increase the number of detectable galaxies at
high redshift by changing the detection criteria and/or by changing the
redshift dependence of disk sizes.  In this case, however, too many
faint galaxies would emerge to reconcile with the number of galaxies at
low redshift and with the observed galaxy number counts.  It is evident
that suppression of galaxy formation is necessary only at low redshift
in order to match the redshift distribution for the WFB model with the
observation.  Thus, unless such mechanisms are found, the WFB model is
inconsistent with the observation.

\begin{figure}
%\epsscale{0.7}
\plotone{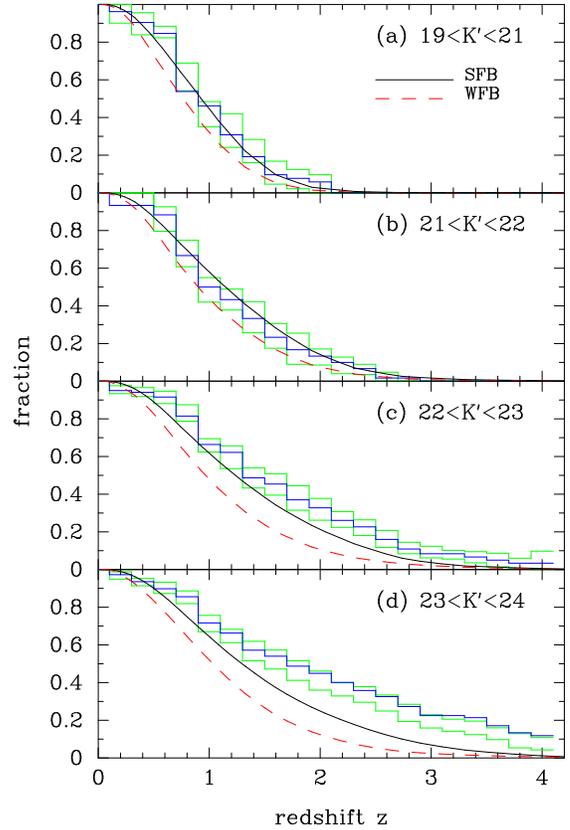}

\caption{Normalized cumulative redshift distribution for the SDF
galaxies for (a) $19<K'\leq 21$, (b) $21<K'\leq 22$, (c) $22<K'\leq 23$,
and (d) $23<K'\leq 24$.  The solid and dashed curves represent the SFB
and WFB models, respectively.  The histograms shown by the solid lines
indicate the observed redshift distribution.  Those indicated by the
thin lines show the $\pm 3 \sigma$ deviated counts estimated by Monte
Carlo realizations when photometric redshift errors are taken into
account [see \citet{k03} for details].}

\label{fig:sds}
\end{figure}

Figure \ref{fig:sds} shows $K'$ band normalized cumulative redshift
distribution for the SDF galaxies.  The solid and dashed lines represent
the SFB and WFB models, respectively.  Only models with the selection
effects are plotted.  The histograms indicated by the solid lines are
the observed redshift distribution taken from \citet{k03}.  Those
indicated by the thin lines show the $\pm 3 \sigma$ deviated counts
estimated by Monte Carlo realizations when photometric redshift errors
are taken into account.  The SFB model agrees significantly well with
the observation in the top two panels {\it a} and {\it b}, which shows
galaxies in the two brightest bins.  In panel {\it c}, the agreement is
only marginal.  Although in panel {\it d}, the deviation from the
observation is large, we should note quite large errors in estimating
photometric redshift in this magnitude range, as shown in Figure 2 of
\citet{k03}.  On the other hand, the difference between the WFB model
and the observation is large for faint galaxies, $K'>22$ (panels {\it c}
and {\it d}).  The main epoch of galaxy formation in the WFB model is at
lower redshift than that in the SFB model, as shown in Figure
\ref{fig:z}.

In \citet{k03}, the observational data were compared with an old version
of the Mitaka model given by \citet{nytg02}.  In that model, parameters
are similar to those of the SFB model, rather than those of the WFB
model.  Nevertheless, the behavior of the old Mitaka model is very
similar to that of the WFB model.  We have confirmed that the Mitaka
model presented in NY04 also shows a tendency similar to the WFB model.
The reason the redshift distribution for the SFB model is improved
should be the adoption of merging histories of dark halos directly taken
from $N$-body simulations.  This might suggest that using correct
merging histories is crucial to understanding of high redshift galaxies.

\begin{figure}
%\epsscale{0.7}
\plotone{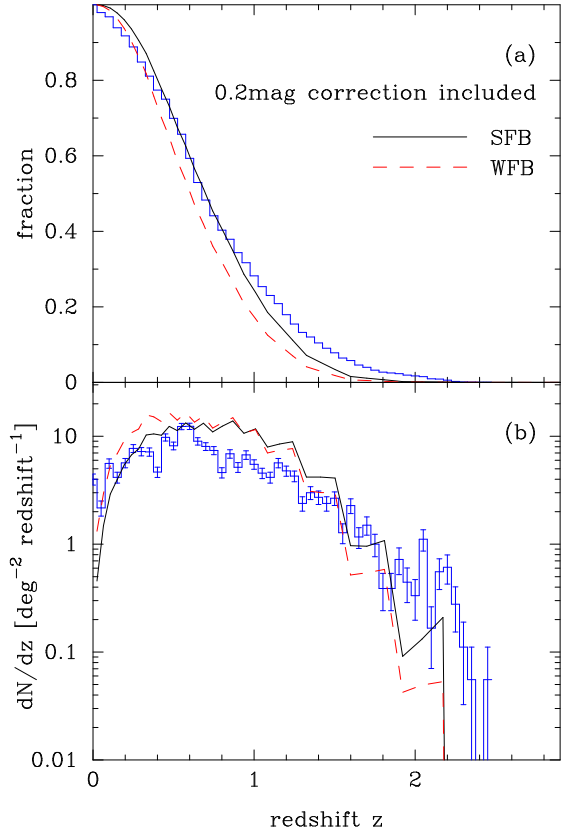}

\caption{(a) Normalized cumulative redshift distribution for the GOODS
$K_{AB}<22$ galaxies.  (b) Same as panel {\it a} but for the
differential redshift distribution.  The solid and dashed lines
represent the SFB and WFB models, respectively.  The histograms are the
observed redshift distribution.  Error bars in panel {\it b} denote 1
$\sigma$ Poisson scatter.  To obtain the isophotal magnitude used in the
GOODS observations, a correction of 0.2 mag is applied to the models
according to \citet{somerville04}.}

\label{fig:GOODS1}
\end{figure}

Finally, we compare our models with recent, larger samples from the
Great Observatories Origins Deep Survey (GOODS) data.  Figure
\ref{fig:GOODS1} shows the (a) normalized cumulative redshift
distribution and (b) differential redshift distribution for the GOODS
galaxies of $K_{AB}(=K_{\rm Vega}+1.85)<22$.  The solid and dashed
curves represent the SFB and WFB models, respectively.  The histograms
are the observed GOODS redshift distribution from \citet{somerville04}.
Error bars in panel {\it b} denote 1 $\sigma$ Poisson scatter.  To
obtain the isophotal magnitude, according to their paper, we make all
galaxies 0.2 mag fainter than their total magnitudes.  The SFB model
seems to provide redshift distribution similar to their SA model shown
in their Figure 1.  Although the SFB model a little overestimates the
number of galaxies at $z\sim 1$, the predicted number of those galaxies
at $z\ga 1.5$ seems slightly closer to the observation than their SA
model.  In the WFB model, similar to the previous figures on redshift
distribution, the agreement with the observation is worse compared with
the SFB model and the fraction of low redshift galaxies becomes larger.
This is also because of the weakness of the SN feedback.  Thus, we can
also say that the SFB model is more consistent with the GOODS redshift
distribution than the WFB model.

In short, the SFB model is consistent with the photometric redshift
distribution for $I_{814}<28$ and $K'<23$ and the WFB only for
$I_{814}<24$ and $K'<22$.  Currently, observational data are still
limited in number and in accuracy of estimation of redshift,
particularly for near-infrared pass-bands.  Wider and deeper surveys in
future with accurate determination of redshifts, will provide better
constraints on galaxy formation.

\subsection{Angular sizes}

\begin{figure}
\plotone{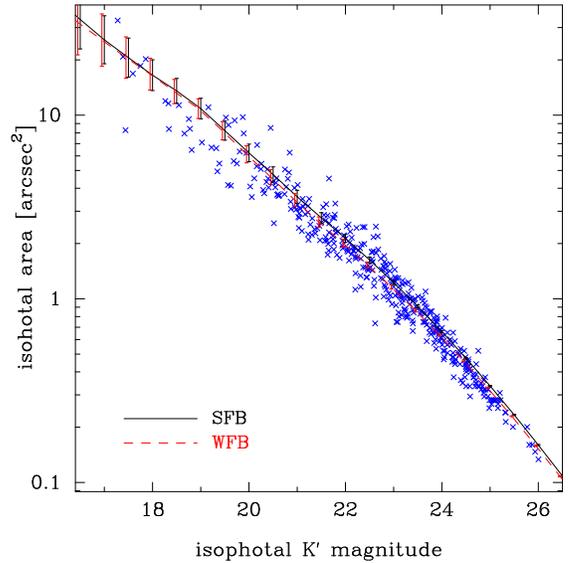}

\caption{Isophotal area of the SDF galaxies against $K'$-magnitude.  The
solid and dashed lines associated with error bars of 1 $\sigma$ scatter
represent the SFB and WFB models, respectively.  To avoid confusion,
small offsets are added to the magnitude of the WFB model.  The isophoto
is calculated from the observational condition employed in the SDF
survey.  The crosses indicate the observational data from the SDF
survey.}

\label{fig:angsize}
\end{figure}

Figure \ref{fig:angsize} shows the isophotal area of galaxies plotted
against their $K'$-magnitude.  The solid and dashed lines associated
with error bars of 1 $\sigma$ scatter represent the SFB and WFB models,
respectively.  To avoid confusion, small offsets are added to the
magnitude of the WFB model.  To obtain the isophoto, the same
observational condition employed in the SDF survey is used.  The crosses
indicate the observational data from the SDF survey.  This good
agreement with the observations means that the models correctly estimate
not only the size of galaxies, but the selection effects arising from
the cosmological dimming of surface brightness of galaxies.

\subsection{Colors}\label{sec:color}

The evolution of colors of galaxies depends on growth of stellar mass,
stellar ages, and dust extinction.  Thus we are able to extract a great
deal of information from galactic colors.

\begin{figure*}
\plotone{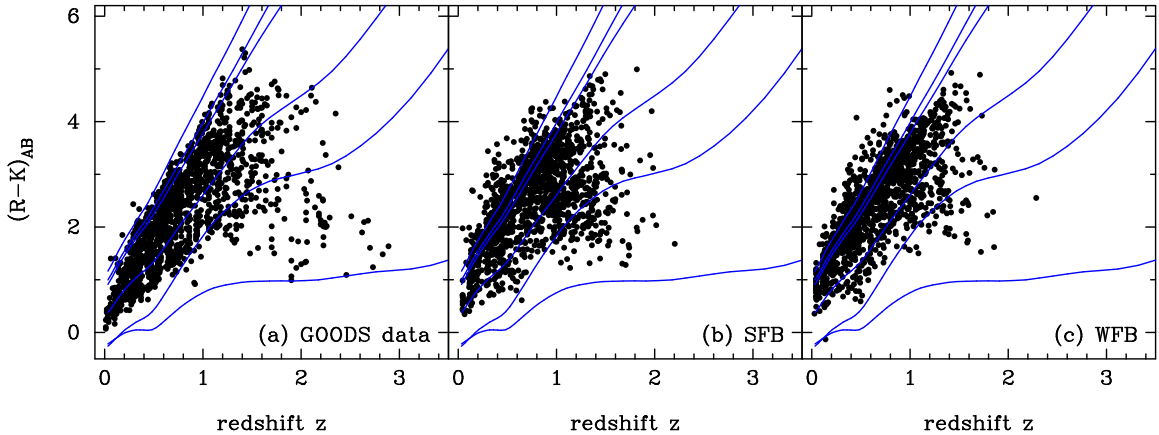}

\caption{Color-redshift diagram for the (a) GOODS galaxies, (b) SFB
model, and (c) WFB model.  The color $(R-K)_{AB}$ is defined in the
observer frame.  The observed data are taken from Figure 2 in
\citet{somerville04}.  In panels {\it b} and {\it c}, we randomly pick
out galaxies of $K_{AB}<22-0.2$, where 0.2 means the correction to
obtain the isophotal magnitude, from the mock catalogs based on the SFB
and WFB models in order to be a number similar to the GOODS samples.
The solid lines indicate evolutionary tracks of single stellar
populations of solar metallicity with ages of 13.5, 5.8, 3.2, 1, 0.5,
and 0.1 Gyr, from the top to the bottom, computed using the model given
by \citet{ka97}.  These ages are the same as those in Figure 2 of
\citet{somerville04}, selected for ease of comparison.}

\label{fig:RK}
\end{figure*}

Figure \ref{fig:RK} shows $R-K$ colors of galaxies in the AB system
against their redshift.  Because of comparison with the GOODS galaxies,
we also adopt the selection criterion $K_{AB}+0.2<22$, where we add 0.2
mag to our models to obtain the isophotal magnitude.  In panel {\it a},
the GOODS galaxies, retrieved from Figure 2 of \citet{somerville04}, are
plotted.  In panels {\it b} and {\it c}, galaxies in the SFB and WFB
models are plotted, respectively.  Among galaxies satisfying the above
criterion, we randomly pick out galaxies so as to show almost the same
number of galaxies as that of the GOODS samples.  The solid lines
indicate evolutionary tracks of single stellar populations of the solar
metallicity with ages of 13.5, 5.8, 3.2, 1, 0.5, and 0.1 Gyr, from the
top to the bottom, estimated using the model given by \citet{ka97}.
These ages are the same as those in Figure 2 of \citet{somerville04} so
that we can easily compare our results with theirs.  As can be seen,
both the SFB and WFB models are in good agreement with the GOODS survey.
\citet{somerville04} has concluded that their SA model broadly agrees
with the GOODS survey, while a small discrepancy remains insofar as
their SA model does not produce enough galaxies with very red colors,
$(R-K)_{AB}\ga 4$, at $z\sim 0.8-1.2$.  Our models seem to be able to
produce more red galaxies, but slight discrepancies also remain.  As
shown below, one of these is the lack of galaxies with $(R-K)_{AB}\sim
2$ at $z\sim 2.5$, and another is the existence of too many galaxies
above the 13.5 Gyr single stellar population line.  A key to
understanding the evolution of colors should be dust extinction.  Our SA
model has two types of dust extinction associated with the two modes of
star formation, quiescent disk star formation and starbursts.  By
switching on and off the dust extinction, we investigate the effects of
dust on colors.

\begin{figure*}
\plotone{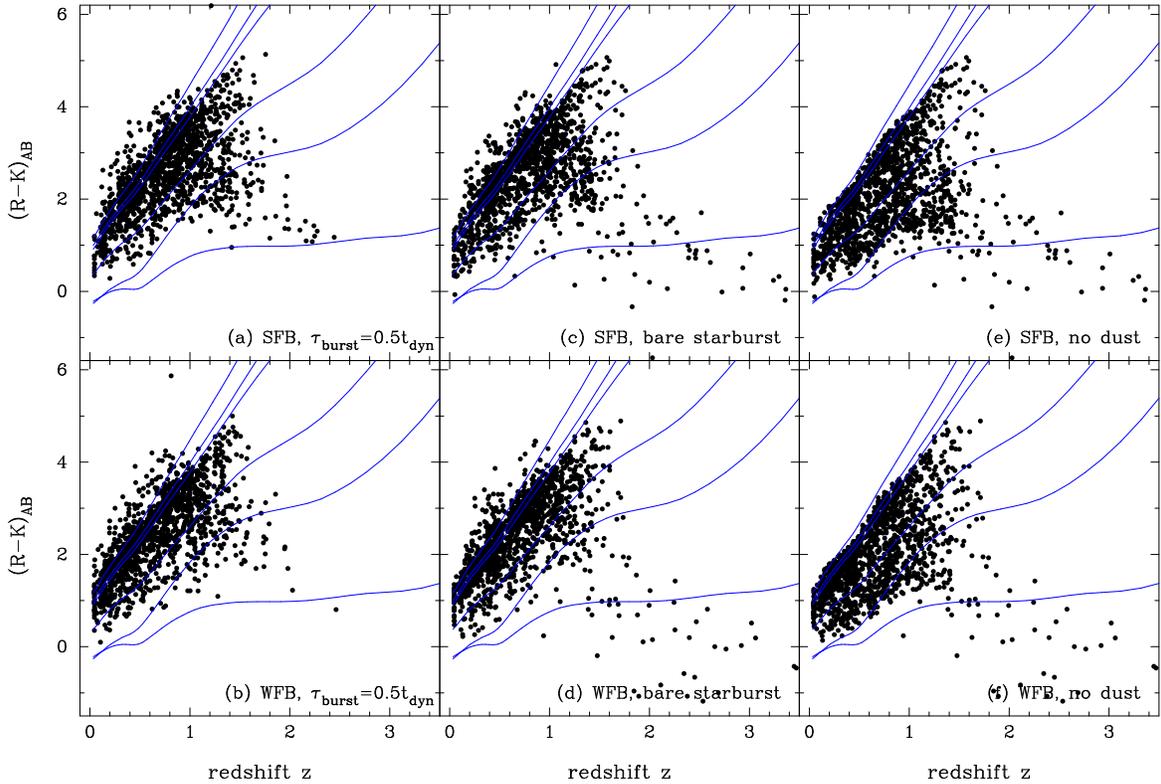}

\caption{Color-redshift diagram for the (a) SFB and (b) WFB models with
$\tau_{\rm burst}=0.5t_{\rm dyn}$, the (c) SFB and (d) WFB models with
no dust extinction during starbursts, and the (e) SFB and (f) WFB models
with no dust extinction at all.  The solid lines are the same as those
in Figure \ref{fig:RK}.}

\label{fig:RK2}
\end{figure*}

Here we examine the effects of dust extinction for starbursts.  As
explained in \S\S \ref{sec:photo}, the SF time-scale for starbursts,
$\tau_{\rm burst}$, is assumed to be equal to the dynamical time-scale
of the spheroidal component, $t_{\rm dyn}\equiv r_{e}/V_{b}$.  Firstly
we change this relation to $\tau_{\rm burst}=0.5t_{\rm dyn}$, that is,
we take half the dynamical time-scale as the SF time-scale for
starbursts.  In panels {\it a} and {\it b} of Figure \ref{fig:RK2}, we
show the same as Figure \ref{fig:RK}, but for the above short starburst
time-scale for the SFB and WFB models, respectively.  It is evident
that, in both models, many galaxies with $(R-K)_{AB}\sim 2$ at $z\sim 2$
emerge and therefore the agreement with the observation is much
improved, though not perfect.  The reason for this improvement is
explained as follows: The time-scale of starbursts in our SA model is
distributed in a broad range comparable to the time-step, about 0.3-0.5
Gyr, at low redshift.  Right after starbursts, the age effect on the
luminosity, even for such a short time-scale, is significant, so we put
the epoch of starbursts within the time-step randomly, depending on the
merger time-scale.  Therefore, for some galaxies, starbursts happen just
before the output redshift.  Such galaxies just after starbursts are
expected to be intrinsically very blue.  When the SF time-scale for
starbursts becomes shorter, the amount of dust estimated from the amount
of cold gas and metals becomes smaller at the output redshift.  Thus, by
shortening the SF time-scale for starbursts, several blue galaxies
emerge.

This can be clearly seen if we do not include the dust extinction for
starbursts at all.  In the center panels {\it c} and {\it d} of Figure
\ref{fig:RK2}, we show the same models as the SFB and WFB models but
without dust extinction during starburst phases (`bare' starbursts).  At
high redshift $z\ga 1$, very blue galaxies with $(R-K)_{AB}\la 0$ emerge
in both the SFB and WFB models.  These galaxies are those just after
starbursts.  The GOODS data at such a high redshift suggest that clearly
starbursts should not be bare, but our dust model for starbursts might
be too strong.  We have also tried the slab dust model instead of the
screen dust model, but found that it does not work well.  It should also
be noted that dust extinction can remove galaxies from this figure
because galaxies do not satisfy the selection criterion,
$K_{AB}+0.2<22$.  More sophisticated modeling of dusty starbursts would
be required in future.

Next, we discuss the extinction of disk luminosities.  In the right
panels {\it e} and {\it f} of Figure \ref{fig:RK2}, we show the same
models as the SFB and WFB models, but without any dust extinction.  By
this manipulation, galaxies above the line of single stellar populations
of 13.5 Gyr disappear, similar to the GOODS data.  This might suggest
that our dust model for disks is stronger than the observed galaxies.
Part of the reason should be the higher value of the chemical yield we
use, because the optical depth is proportional to the metallicity of
cold gas, as mentioned in \S\S \ref{sec:cmr}.  Presumably, careful
choice of IMFs, which determine the chemical yield, depending on the SF
mode will improve the color distribution.

It should be noted that the changing of the optical depth for disks may
affect various observables.  For example, weakening the dust extinction
in a simple way would cause overestimation of faint galaxy counts in
optical pass-bands.  In order to obtain full agreement after changing
the optical depth, therefore, a new set of model parameters will be
required.  Instead, if the extinction only for strongly reddened
galaxies is weakened, it might not significantly affect model
parameters.  The change of $\gamma$ might also recover the agreement
with the observational galaxy counts.  In all cases, the dust model
adopted in this paper is very simple and should be refined in future.
By contrast, the dust model for starbursts is not the case because the
number fraction of bursting galaxies is smaller than that of normal
galaxies.  As shown in the next subsection, however, it might affect the
number of high redshift galaxies at $z\ga 3$, at which a substantial
fraction of stars is formed via starbursts, while the contribution to
faint galaxy counts is small.

\subsection{Cosmic Star Formation Histories}

\begin{figure}
\plotone{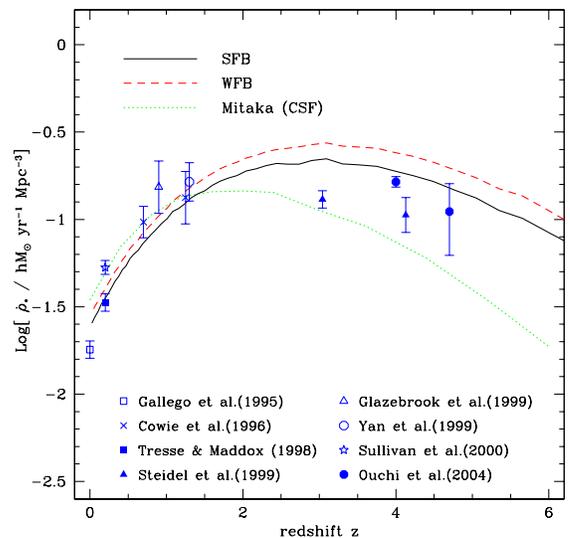}

\caption{Cosmic SF histories.  The solid and dashed lines represent the
SFB and WFB models, respectively.  For reference, the Mitaka CSF model
in NY04 is also plotted by the dotted line.  Symbols with error bars are
the observed cosmic SFRs as indicated in the figure.}

\label{fig:csfr1}
\end{figure}

Recently the cosmic SF history, which is a plot of SFRs in a comoving
volume against redshift, has been recognized as one of the fundamental
quantities characterizing galaxy formation \citep{madau96}, in spite of
large uncertainties in its estimation caused by, for example, dust
obscuration and shape of the luminosity function below the detection
limit.  Figure \ref{fig:csfr1} shows the cosmic SFR as a function of
redshift.  The solid and dashed lines represent the SFB and WFB models,
respectively.  For comparison, the Mitaka CSF model, in which the SF
time-scale is independent of redshift similar to the SFB and WFB models,
is also plotted by the dotted line.  Symbols are the observational data
from surveys indicated in the figure.  Note that the observational
values are corrected, taking into account dust extinction, and also that
there are still large uncertainties originating in undetected faint
galaxies at high redshift.  As shown in Figure 21 of \citet{ouchi04},
observationally obtained cosmic SFRs strongly depend on the assumed
luminosity function, which is guessed from the data for detected bright
galaxies.  The upper limits for values at $z\ga 3$ estimated by
\citet{ouchi04} are about a factor of two larger than plotted values.
These are obtained by integrating luminosity functions down to the
luminosity $L=0$.  Taking into account such uncertainties which are not
included in plotted error bars, the SFB and WFB models are in good
agreement with the observations.

As we have already shown in NY04, the cosmic SFR strongly depends on the
model of the SF time-scale.  If the SF time-scale becomes shorter toward
higher redshift, for example, proportional to the dynamical time-scale
of disks or halos, the peak of the cosmic SFR moves to a higher
redshift.  Because the SFB and WFB models have the same SF model, which
is independent of redshift, their predicted cosmic SFRs are similar.
The Mitaka CSF model, however, shows different cosmic SFRs, in spite of
having the same redshift dependence as the SF time-scale.  This would
suggest that it is very important for high redshift galaxies to
correctly estimate merging histories of dark halos, as pointed out in
\S\S \ref{sec:zdist}.

\begin{figure}
\plotone{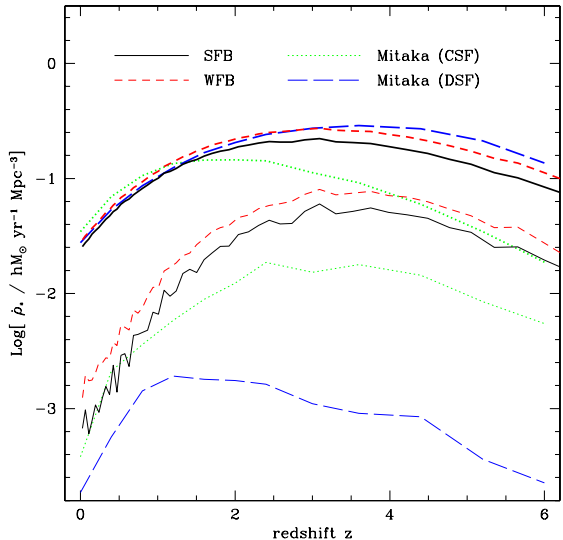}

\caption{Cosmic SF histories.  The solid and short dashed lines
represent the SFB and WFB models, respectively.  The dotted and long
dashed lines represent the Mitaka CSF and DSF models, respectively, in
NY04.  The thick lines indicate the total SFR, and the thin lines the
SFR only for starbursts.}

\label{fig:csfr2}
\end{figure}

Figure \ref{fig:csfr2} shows not only the total cosmic SFR ({\it thick
lines}) but also the SFR only for starbursts ({\it thin lines}).  The
solid and short dashed lines represent the SFB and WFB models,
respectively.  The dotted and long dashed lines represent the Mitaka CSF
and DSF models, respectively.  The former has a SF time-scale
independent of redshift, and the latter a SF time-scale proportional to
the dynamical time-scale of dark halos, which becomes shorter toward
higher redshift.  Although the total cosmic SFRs of the SFB and WFB
models are similar to the Mitaka DSF model rather than the Mitaka CSF
model, SFRs for starbursts are quite different.  The SFB and WFB models
predict a very large contribution of starbursts to the total SFR up to
15-20\% at $z\sim 2$ and 25-30\% at $z\sim 4$, while the Mitaka CSF
model predicts a contribution of starbursts about 10\% at $z\sim 2$ and
20\% at $z\sim 4$ and the Mitaka DSF model at most 2\% at $z\sim 1$.  If
the IMF for starbursts is top-heavy, as suggested by \citet{baugh05} and
\citet{nlbfc05}, this significant fraction would affect many aspects,
such as the chemical enrichment and the cosmic rates of supernovae and
$\gamma$-ray bursts.

\section{Summary}

We have developed a semi-analytic (SA) model combined with high
resolution $N$-body simulations, which we call the {\it Numerical Galaxy
Catalog} ($\nu$GC).  The minimum mass of dark halos is down to
3.04$\times 10^{9} M_{\odot}$, which corresponds to $V_{\rm circ}\sim
20$ km s$^{-1}$ at $z=0$ and 40 km s$^{-1}$ at $z\simeq 4$.  These are
enough to resolve the effective Jeans mass, or the filtering mass, after
the cosmic reionization.  To examine the effects of the resolution on
galaxy formation, we also used another simulation with a box twice the
size, in which the minimum mass of dark halos is $2.43\times 10^{10}
M_{\odot}$ corresponding to $V_{\rm circ}\sim 40$ km s$^{-1}$ and 80 km
s$^{-1}$.  Although this resolution is not significantly worse among the
currently available cosmological $N$-body simulations, we have found
that it alters luminosity functions, especially the faint-end slope.
Because SN feedback-related parameters ($\alpha_{\rm hot}$ and $V_{\rm
hot}$) are determined so as to match luminosity functions, worse
resolutions should affect values of parameters.  We would like to stress
that the resolution of our SA model is much better than that of
currently available data provided by SA models combined with $N$-body
simulations in the public domain.  For example, the particle mass of
dark matter in the GIF project \citep{kcdw99a} is about 65 times larger
than ours, and that of the GalICS project \citep{hatton03} is about 27
times larger.  These are even worse than our 140 $h^{-1}$ Mpc box
simulation.

Within observational uncertainties of the shape of luminosity functions,
we have adopted two parameter sets.  One is the strong SN feedback (SFB)
model and another is the weak SN feedback (WFB) model.  In this paper,
we have focused on comparison of these models with photometric,
structural, and kinematical properties of observed galaxies.  Both
models reproduce many observations well, and we slightly prefer the SFB
model to the WFB model.  Below, we summarize the results.
\\

 1.  We have explicitly shown that both the SFB and WFB models are in
 reasonable agreement with the following observational results:
 luminosity functions in $B$ and $K$ bands, cold gas mass-to-$B$ band
 luminosity ratios of spiral galaxies, \ion{H}{1} mass functions,
 size-magnitude relations for spiral and elliptical galaxies, and the
 Faber-Jackson relation at $z=0$, and faint galaxy number counts in
 $BVRi'z'$ bands, isophotal area, and cosmic star formation histories at
 high redshift.  Note that some of the above observations were used to
 fix the model parameters.  For example, the SN feedback-related
 parameters were determined by matching the luminosity functions.  Table
 \ref{tab:astro} summarizes observations to fix the parameters.
\\

 2.  The SFB model is in broad agreement with the observed galaxy counts
 in the near-infrared $K'$ band.  Most observed redshift distributions
 also are broadly reproduced by the SFB model, except for the faintest
 galaxies ($23<K'\leq 24$) in the Subaru Deep Survey, while the results
 are much improved compared with those of the Mitaka model based on the
 analytic EPS formalism, not on $N$-body simulations.  However, since
 errors caused by uncertainties in estimating photometric redshifts are
 very large for such faint galaxies, this discrepancy may not be taken
 so seriously.  Future observations with accurate determination of
 redshifts and with a large survey area will provide sufficiently better
 redshift distributions for very faint galaxies to make clear whether
 models are in agreement with observations or not.  By contrast, the WFB
 model is not consistent with the observed $K'$ band galaxy counts in
 the Subaru Deep Field, while it is in agreement with optical
 ($BVRi'z'$) galaxy counts.  This model is also inconsistent with
 redshift distributions, in particular, for faint galaxies.
\\

 3.  The Tully-Fisher (TF) relation slightly favors the WFB model to the
 SFB model, because the slope of the TF relation predicted by the SFB
 model is steeper than the observed one.  This contradiction in the TF
 relation may be improved if we introduce dynamical response on the
 rotation velocity to gas removal induced by the SN feedback, which is
 only taken into account for starbursts making spheroidal systems in the
 current model.  We are going to explicitly consider this effect on
 spiral galaxies in future.
\\

 4.  The bright end of color-magnitude relations of cluster elliptical
 galaxies in the SFB and WFB models is almost flat against magnitudes
 and also inconsistent with observations.  Presumably, this will be
 amended by adopting different IMFs for disk star formation and
 starbursts.  Introducing the chemical enrichment due to SNe Ia into the
 Mitaka model, \citet{no04} have shown that chemical yields consistent
 with Salpeter's IMF are in agreement with observed metallicities and
 their ratios in the solar neighborhood.  Moreover, \citet{nlbfc05} have
 also shown that metallicities of the intracluster medium can be
 reproduced only when adopting a top-heavy IMF for starbursts and
 Kennicutt's IMF for disk star formation.  Therefore it is not
 unreasonable to adopt different IMFs.  At this stage, however, it is
 beyond the scope of this paper.
\\

 5.  Both models predict $(R-K)_{AB}$ colors against redshifts broadly
 consistent with the observational GOODS galaxies.  Particularly, our
 models are able to produce red galaxies $(R-K)_{AB}\sim 5$, which have
 not been reproduced by an SA model of the GOODS group
 \citep{somerville04}.  However, slight discrepancies remain.  One is
 lack of galaxies with $(R-K)_{AB}\sim 1-2$ at $z\ga 2$.  If we switch
 off the dust extinction for starburst galaxies, several galaxies
 emerge.  At the same time, however, very blue galaxies with
 $(R-K)_{AB}\la 1$ also emerge.  When a shorter star formation
 time-scale is used for starbursts similar to half the dynamical
 time-scale, the situation can be improved.  This means that more
 sophisticated modeling of dusty starbursts should be required.  Second
 is that our model produces excessively red galaxies at low redshift,
 $z\la 1$.  This is caused by too strong a dust extinction for disk
 stars.  This might suggest that the chemical yield used here, which is
 almost twice the solar metallicity, is too large, because dust optical
 depth is proportional to the metallicity of cold gas.  This high value
 is required to produce red elliptical galaxies to be consistent with
 observed cluster ellipticals.  Therefore, similar to color-magnitude
 relations, adopting different IMFs for disk star formation and
 starbursts will help these galaxies become blue.
\\

This is a first paper in a series of $\nu$GC.  Subsequent papers will
discuss clustering properties of spatial distribution of galaxies
\citep{ynegy}.  We also plan to combine the quasar formation model given
by \citet{eng03} with this model \citep{eyngy}.  $\nu$GC will be a
useful tool with which to understand galaxy formation in studies of
various current and future large galaxy surveys.

\acknowledgments

We would like to thank Takashi Okamoto for useful comments on modeling
galaxy formation with $N$-body simulations, Nobunari Kashikawa for
useful discussion on the Subaru Deep Survey, Bahram Mobasher for
providing the GOODS data of redshift distribution plotted in Figure
\ref{fig:GOODS1}, Masami Ouchi for providing the data of cosmic SFRs
plotted in Figure \ref{fig:csfr1}, and Katsuya Okoshi for useful
discussion on \ion{H}{1} mass functions.  This work has been supported
in part by the Grant-in-Aid for the Center-of-Excellence research
(07CE2002) of the Ministry of Education, Science, Sports and Culture of
Japan.  MN acknowledges support from the PPARC rolling grant for
extragalactic astronomy and cosmology at Durham and from the Japan
Society for the Promotion of Science for Young Scientists (No.207).
$N$-body simulations described in this paper were carried out using
Fujitsu-made vector parallel processors VPP5000 installed at the
Astronomical Data Analysis Center, National Astronomical Observatory,
Japan (ADAC/NAOJ), under the ADAC/NAOJ large scale simulation projects
(group-ID: myy26a, yhy35b).

\end{document}